\documentclass[twocolumn]{aastex63}
\usepackage{newtxtext,newtxmath}
\usepackage[T1]{fontenc}

\DeclareRobustCommand{\VAN}[3]{#2}
\let\VANthebibliography\thebibliography
\def\thebibliography{\DeclareRobustCommand{\VAN}[3]{##3}\VANthebibliography}

\usepackage{graphicx}	% Including figure files
\usepackage{amsmath}	% Advanced maths commands
\usepackage{soul} % use this (many fancier options)
\usepackage{url}
\usepackage{lineno}
\usepackage{comment}

\newcommand{\lenstool}{{\tt{Lenstool}}}

\newcommand{\HST}{{\it HST}}
\newcommand{\JWST}{{\it JWST}}
\newcommand{\jwst}{{\it JWST}}

\newcommand{\review}[1]{#1}

\newcommand{\msun}{M$_{\odot}$}

\newcommand{\rms}{{0\farcs32}} %{\textcolor{red}{0\farcs32}} % RMS of this model
\newcommand{\SMACSlong}{SMACS\,J0723.3$-$7327}
\newcommand{\SMACS}{SMACS\,J0723}

\received{xxx}
\revised{xxx}
\accepted{xxx}%\today

\begin{document}

\title{Precision modeling of JWST's first cluster lens SMACS\,J0723.3$-$7327\footnote{Based on observations made with the NASA/ESA \textit{Hubble Space Telescope}, obtained at the Space Telescope Science Institute, which is operated by the Association of Universities for Research in Astronomy, Inc., under NASA contract NAS 5-26555. These observations are associated with programs GO-11103, GO-12166, GO-12884, GO-1409; and on observations made with the NASA/ESA/CSA \textit{James Webb Space Telescope}, under NASA contract NAS 5-03127 for \textit{JWST}. These observations are associated with program \#2736.}}
\shorttitle{First JWST lensing Cluster SMACS\,J0723.3$-$7327}
\shortauthors{Mahler et al.}

\correspondingauthor{Guillaume Mahler}
\email{guillaume.mahler@durham.ac.uk }

\author[0000-0003-3266-2001]{Guillaume Mahler}
\affiliation{Centre for Extragalactic Astronomy, Durham University, South Road, Durham DH1 3LE, UK}
\affiliation{Institute for Computational Cosmology, Durham University, South Road, Durham DH1 3LE, UK}

\author[0000-0003-1974-8732]{Mathilde Jauzac}
\affiliation{Centre for Extragalactic Astronomy, Durham University, South Road, Durham DH1 3LE, UK}
\affiliation{Institute for Computational Cosmology, Durham University, South Road, Durham DH1 3LE, UK}
\affiliation{Astrophysics Research Centre, University of KwaZulu-Natal, Westville Campus, Durban 4041, South Africa}
\affiliation{School of Mathematics, Statistics \& Computer Science, University of KwaZulu-Natal, Westville Campus, Durban 4041, South Africa}

\author[0000-0001-5492-1049]{Johan Richard}
\affiliation{Univ Lyon, Univ Lyon1, Ens de Lyon, CNRS, Centre de Recherche Astrophysique de Lyon UMR5574, 69230, Saint-Genis-Laval, FR}

\author[0000-0002-0443-6018]{Benjamin Beauchesne}
\affiliation{Institute of Physics, Laboratory of Astrophysics, École Polytechnique Fédérale de Lausanne (EPFL), Observatoire de Sauverny, CH-1290 Versoix, Switzerland}
\affiliation{European Southern Observatory, Alonso de Cordova 3107, Vitacura, Santiago, Chile}

%Harald Ebeling$^{27}$,
\author[0000-0001-8249-2739]{Harald Ebeling} 
\affiliation{Institute for Astronomy, University of Hawaii, 640 N Aohoku Pl, Hilo, HI 96720, USA}

%David Lagattuta$^{x}$,
\author[0000-0002-7633-2883]{David Lagattuta} 
\affiliation{Centre for Extragalactic Astronomy, Durham University, South Road, Durham DH1 3LE, UK}
\affiliation{Institute for Computational Cosmology, Durham University, South Road, Durham DH1 3LE, UK}

\author[0000-0002-5554-8896]{Priyamvada Natarajan} \affiliation{Department of Astronomy, Yale University, 52 Hillhouse Avenue, New Haven, CT 06520, USA}
\affiliation{Department of Physics, Yale University, P.O. Box 208121, New Haven, CT 06520, USA}
\affiliation{Black Hole Initiative, Harvard University, 20 Garden Street, Cambridge MA 02138, USA} 

\author[0000-0002-7559-0864]{Keren Sharon}
\affiliation{Department of Astronomy, University of Michigan, 1085 S. University Ave, Ann Arbor, MI 48109, USA}

%Hakim Atek$^{x}$, 
\author[0000-0002-7570-0824]{Hakim Atek} 
\affiliation{Institut d’astrophysique de Paris, CNRS, Sorbonne Universite, 98bis Boulevrad Arago, 75014, Paris, France}

\author[0000-0001-7940-1816]{Adélaïde Claeyssens}
\affiliation{The Oskar Klein Centre, Department of Astronomy, Stockholm University, AlbaNova, SE-10691 Stockholm, Sweden}

%Benjamin Clément$^{3}$, 
\author[0000-0002-7966-3661]{Benjamin Clément}
\affiliation{Institute of Physics, Laboratory of Astrophysics, École Polytechnique Fédérale de Lausanne (EPFL), Observatoire de Sauverny, CH-1290 Versoix, Switzerland}

%and Dominique Eckert$^{y}$
\author[0000-0001-7917-3892]{Dominique Eckert} 
\affiliation{Department of Astronomy, University of Geneva, ch. d'Ecogia 16, CH-1290 Versoix, Switzerland}

%Alastair Edge$^{x}$, 
\author[0000-0002-3398-6916]{Alastair Edge} 
\affiliation{Centre for Extragalactic Astronomy, Durham University, South Road, Durham DH1 3LE, UK}

%Jean-Paul Kneib$^{3}$,
\author[0000-0002-4616-4989]{Jean-Paul Kneib} 
\affiliation{Institute of Physics, Laboratory of Astrophysics, École Polytechnique Fédérale de Lausanne (EPFL), Observatoire de Sauverny, CH-1290 Versoix, Switzerland}

\author[0000-0003-3791-2647]{Anna Niemiec}
\affiliation{Centre for Extragalactic Astronomy, Durham University, South Road, Durham DH1 3LE, UK}
\affiliation{Institute for Computational Cosmology, Durham University, South Road, Durham DH1 3LE, UK}

% Abstract of the paper
\begin{abstract}
Exploiting the fundamentally achromatic nature of gravitational lensing, we present a lens model for the massive galaxy cluster SMACS\,J0723.3$-$7323 (SMACS\,J0723, $z=0.388$) that significantly improves upon earlier work. Building on strong-lensing constraints identified in prior \emph{Hubble Space Telescope} (\emph{HST}) observations, the mass model utilizes 21 multiple-image systems, 17 of which were newly discovered in Early Release Observation (ERO) data from the \emph{James Webb Space Telescope} (\emph{JWST}). The resulting lens model maps the cluster mass distribution to an RMS spatial precision of \rms\ and is publicly available\footnote{\url{https://github.com/guillaumemahler/SMACS0723-mahler2022}}. Consistent with previous analyses, our study shows SMACS\,J0723.3$-$7323 to be well described by a single large-scale component centered on the location of the brightest cluster galaxy. However, satisfying all lensing constraints provided by the JWST data, the model point to the need for the inclusion of an additional, diffuse component west of the cluster. A comparison of the galaxy, mass, and gas distributions in the core of SMACS\,J0723 based on \HST, \JWST, and {\it Chandra} data reveals a concentrated regular elliptical profile along with tell-tale signs of a recent merger, possibly proceeding almost along our line of sight. The exquisite sensitivity of \JWST's NIRCAM reveals in spectacular fashion both the extended intra-cluster-light distribution and numerous star-forming clumps in magnified background galaxies. The high-precision lens model derived here for SMACS\,J0723$-$7323 demonstrates the unprecedented power of combining \HST\ and \JWST\ data for studies of structure formation and evolution in the distant Universe.
\end{abstract}

%%%%%%%%%%%%%%%%%%%%%%%%%%%%%%%%%%%%%%%%%%%%%%%%%%

%%%%%%%%%%%%%%%%% BODY OF PAPER %%%%%%%%%%%%%%%%%%

\section{Introduction}

Clusters of galaxies grow and evolve through large-scale merging processes and offer many valuable observables for astrophysical and cosmological studies of our Universe. In statistically representative samples, clusters uniquely constrain key parameters of complex physical processes, such as structure formation, but also the cosmological parameters characterizing the underlying world model \citep{Jullo2010,Caminha2017,Schwinn2017,Acebron2017}. By measuring the mass distribution within clusters, we also gain insight into cluster-specific properties, such as their dark-matter content; the detailed spatial distribution and clustering of dark matter; the cluster's merger geometry and history \citep[e.g.,][]{Bradac2008,Umetsu2009,KneibPnat2011,Ebeling2017}. Furthermore, potential offsets between the location of baryonic and dark-matter profiles have been used to probe the nature of dark matter (e.g., its self-interaction cross-section, \citealt{Markevitch2004,Randall2008,Wittman2018,Harvey2019}). 

Strong gravitational lensing provides an observational measure of the total enclosed mass of a cluster at a given radius and thus offers a powerful tool for studying both their dark and luminous matter content. Lensing occurs when the presence and concentration of mass generates a large enough curvature in space–time near the cluster center to make different light paths from the same distant source converge within the field of view of the observer. Since the first spectroscopic confirmation of a giant gravitational arc in Abell\,370 \citep{Soucail1987}, strong gravitational lensing has evolved into a valuable and powerful technique for measuring the total mass of clusters over a wide range of evolutionary states and redshifts \citep[e.g.,][]{Limousin2007,Richard2011,Sharon2015}. 

By refining the mass model of a lensing cluster through the identification of strong-lensing features, it is possible to quantify the magnifying power of the cluster for background sources at a given redshift, thereby calibrating galaxy clusters as cosmic telescopes for studies of the high-redshift Universe (e.g. \citealt{Mahler2019,Fox2022}). The correct identification of multiply imaged background sources is crucial in this context, because these are the principal constraining features that permit us to precisely map the mass distribution in the cluster core. The high angular resolution of the \textit{Hubble Space Telescope} (\HST) has proven invaluable for such work, as determination of the source morphology is instrumental to the task of properly matching multiple lensed images of the same source. 

The most ambitious example of this quest to date was the \textit{Hubble} Frontiers Field Initiative \citep[HFF,][]{Lotz2017} which provided very deep \textit{HST} observations ($\sim$180 orbits per target) in seven optical and near-infrared pass-bands. The HFF observed six massive clusters ($M\approx10^{15}$ \msun) at $z=0.3-0.6$, selected for their lensing power, and, specifically, their capability to strongly magnify very distant ($z>6$) galaxies. The resulting deep images revealed a remarkable collection of hundreds of multiple images that provided unprecedented insights into the detailed mass distribution of clusters and given their visual power were showcased in numerous publications \citep[e.g.,][]{Jauzac2014,Grillo2015,Sharon2015a,Jauzac2016,Diego2016a,Diego2016b,Caminha2017,Mahler2018,Vanzella2021}

\begin{figure*}
	\begin{center}
	    
	\includegraphics[width=\textwidth]{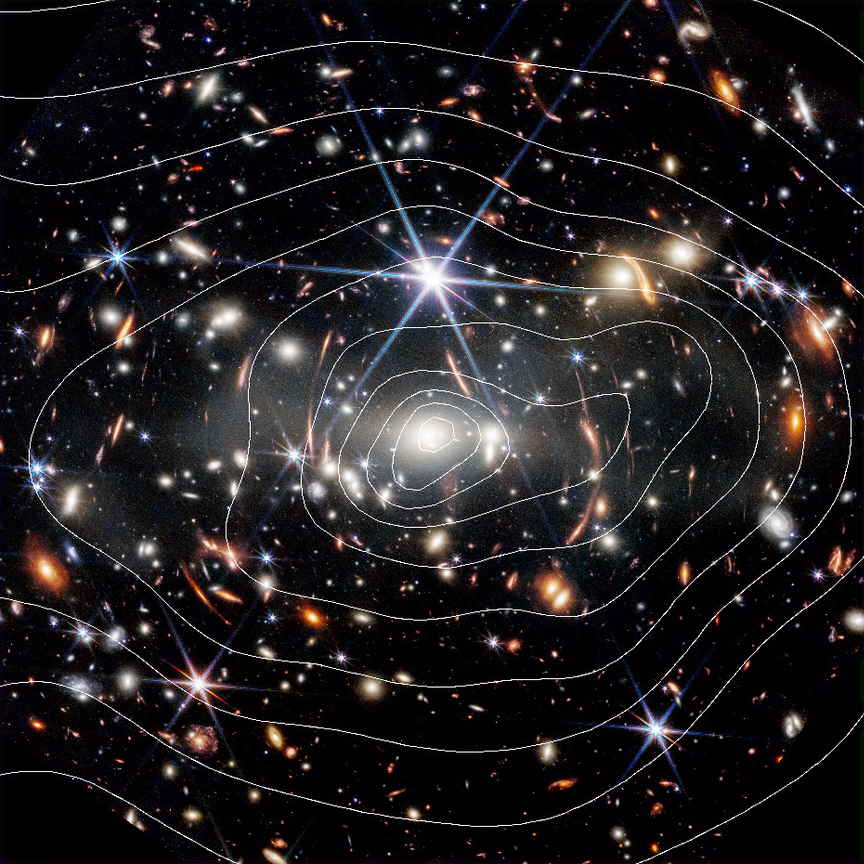}
    \caption{\JWST\ / NIRCam image of a $2\times 2$ arcmin$^2$ area centred on the brightest cluster galaxy (BCG) of SMACS\,J0723. The overlaid white contours show the X-ray surface brightness \citep[adaptively smoothed to $3\sigma$ significance using the algorithm of ][]{2006MNRAS.368...65E} as observed with \textit{Chandra}. Contours are spaced logarithmically by factors of 1.5, starting at three times the background level. The astrometric alignment of the two underlying images is accurate to about 1\arcsec.  %The overlaid magenta contours show the mass density distribution  of the derive lensing mass profile and show remarkable agreement with the X-ray.  In addition, the excess of mass on the right part of the cluster can also be hinted from the smooth low-surface-brightness intra-cluster light (ICL).
    }
    	\end{center}

    \label{fig:xopt}
\end{figure*}

Providing an order-of-magnitude improvement in sensitivity, the \emph{James Webb Space Telescope} (\jwst) represents another dramatic leap forward in our efforts to probe the distant Universe to ever larger depth exploiting gravitational lensing. The enormous promise of \jwst\ is exemplified in the release of \jwst's first deep cluster observation, of \SMACSlong\ (hereafter \SMACS),  results from which are discussed and presented in this paper.

Our paper is structured as follows. After a brief introduction of the target and the history of its discovery (Section~\ref{sec:target}), we summarize the most relevant ground- and space-based observations of SMACS\,J0723 in Section~\ref{sec:obs}. Section~\ref{sec:methods} provides an overview of the analysis and modeling techniques used here, and Section~\ref{sec:results} describes the results obtained \review{from the analysis of \JWST\ data in combination with prior \HST\ data.} We present a summary of our findings and conclusions in Section~\ref{sec:conc}.

For the underlying cosmological model, we assume the $\Lambda$CDM concordance cosmology ($\Omega_{\Lambda} = 0.7$, $\Omega_{m}=0.3$) and $h=0.7$ throughout. In this cosmology, 1\arcsec\ corresponds to 5.3\,kpc at the cluster redshift of $z=0.3877$.

\section{SMACS\,J0723}
\label{sec:target}

SMACS\,J0723 was discovered in the course of the southern extension of the Massive Cluster Survey \citep[MACS;][]{2001ApJ...553..668E} and is included in the partial release of the MACS sample by \citet[][]{2018MNRAS.479..844R}. 

The system's initial identification as a putative distant cluster was based on the presence of an unidentified X-ray source, 1RXS\,J072319.7$-$732735, with 64 detected photons in a 531\,s exposure accumulated during the ROSAT All-Sky Survey \citep[RASS;][]{1999A&A...349..389V}. The source's high X-ray hardness ratio of 0.95 (HR1 in RASS parlance), very high even at the relatively high neutral-hydrogen density of more than $10^{21}$\,cm$^{-2}$ at the source's low Galactic latitude ($b=-23$\,deg), its high likelihood of being extended, as well as the absence of alternative plausible optical counterparts in shallow, archival Digital Sky Survey images, rendered 1RXS\,J072319.7$-$732735 a prime candidate for follow-up observations. Consequently, SMACS\,J0723 was targeted in imaging and low-resolution spectroscopy observations with the 3.5m New Technology Telescope at the European Southern Observatory (ESO) in 2002 and 2003, respectively, which unambiguously confirmed the system as a massive cluster and established its tentative redshift as $z=0.404$ (see Section~\ref{sec:clustergal} for an improved redshift measurement).

\section{Observations and data reduction}
\label{sec:obs}

\subsection{Optical and NIR imaging}
\subsubsection{James Webb Space Telescope}
\begin{figure*}
    \centering
    \includegraphics[width=\textwidth]{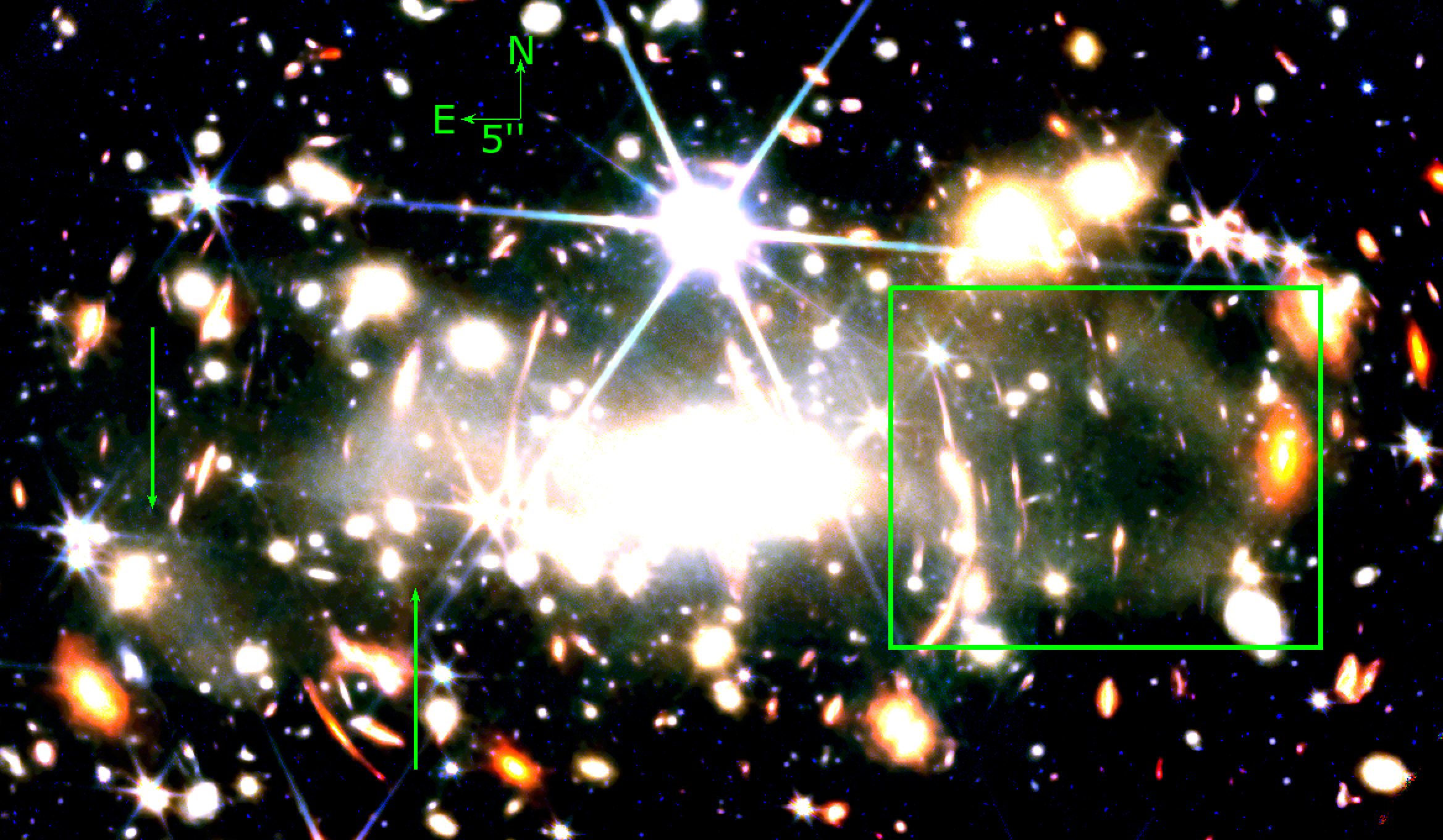}
    \caption{\textit{JWST} / NIRCam image displayed at high contrast after median filtering with a sliding box spanning $21 \times 21$ arcseconds to enhance low-surface-brightness features. Very faint, diffuse emission well beyond the BCG halo is highlighted in a rectangular area west of the cluster core. This image also shows smooth emissions marked by arrows. }
    \label{fig:icl}
\end{figure*}

SMACS\,J0723 was observed in early June 2022 with several instruments aboard \jwst\ as part of the observatory's {\it Early Release Observations}\footnote{\url{https://www.stsci.edu/jwst/science-execution/approved-programs/webb-first-image-observations}}. Specifically, deep imaging was performed in the NIRCAM filters F090W, F150W, F200W, F277W, F356W, and F444W (Figs.~\ref{fig:xopt} and \ref{fig:icl}). The central field was also imaged with MIRI and NIRISS.

Our analysis combines pre-\JWST\ observations (described below) with NIRCAM data and NIRSpec spectroscopy.

\subsubsection{Hubble Space Telescope}
SMACS\,J0723 has been observed several times with multiple instruments aboard the \textit{HST}: first with the Wide Field and Planetary Camera 2 (WFPC2) in the optical regime (F606W and F814W filters) in 2008 (GO-11103, PI Ebeling); then, in the same two filters, with the Advanced Camera for Surveys (ACS) in 2011 and 2014 for GO-12166 and GO-12884, respectively (both PI Ebeling); and finally in 2017 with the Wide-Field Camera 3 (WFC3) and ACS in the F453W, F606W, F814W, F105W, F125W, F140W, and F160W passbands for the RELICS program (GO-14096, PI Coe). In all cases, the observing time ranged from about 1/2 to 1 orbit per filter. Additional snapshot images in the F606W and F105W passbands were obtained with WFC3 in 2022 for GO-16729 (PI Kelly).

\subsection{Spectroscopy}
\label{sec:Spectroscopy}
\subsubsection{ESO}

Shallow (3$\times$970\,s) observations of the cluster were performed in March 2019 in moderate seeing conditions (0.72\arcsec) for Programme 0102.A-0718(A) (PI Edge) with the \review{Multi Unit Spectroscopic Explorer (MUSE)} integral field spectrograph on the ESO Very Large Telescope. The observation covered a 1$\times$1\,arcmin$^{2}$ region centered on the Brightest Cluster Galaxy (BCG) of SMACS\,J0723 and yielded spectra in the range from 480 to 930\,nm of both lensing features and foreground / cluster galaxies.

The reduction of the MUSE data cube was performed using the official ESO pipeline \citep{2020A&A...641A..28W}, with a number of specific improvements regarding self-calibration and sky subtraction specific to the crowded fields of lensing clusters. These are extensively discussed in previously published work by \cite{Lagattuta2022} and \cite{Richard2021}.

\subsubsection{JWST}

As part of the \JWST's Early Release Observations (ERO) of SMACS\,J0723, the observatory also acquired spectroscopic data with the Micro-Shutter-Array (MSA) NIRSpec of 58 individual galaxies, as well as spectra of all objects in the entire field with NIRISS in Wide-Field Slitless mode. The total on-source exposure time ranged from 1.5 to 5\,hrs. Our first analysis presented here uses primarily NIRSpec MSA data (reduced 2D spectra and 1D extracted spectra of various multiple images)  directly available from the ERO data release. 

\subsection{X-ray imaging spectroscopy}

SMACS\,J0723 was observed with the Advanced CCD Imaging Spectrometer (ACIS-I) aboard the \textit{Chandra X-ray Observatory} on April 14, 2014. The observations (Sequence Number 801329; ObsID 15296; PI Murray) were performed in VFAINT mode for a total duration of 19.8\,ks. We performed a standard reduction of the data using the \textsc{CIAO}\footnote{\url{https://cxc.cfa.harvard.edu/ciao/}} $4.13$ \citep{ciao} and \textsc{CALDB} $4.9.6$ packages. We removed point sources detected either automatically by the \textsc{wavedetect} routine or by visual inspection. Periods of background flares were removed by running the \textsc{deflare} tool in the 9.5--12 keV band and for the whole energy range. We used the blank-sky background data associated with the observation as provided by the standard data reduction pipeline.

\section{Methods}
\label{sec:methods}

\subsection{Intra-cluster light}
\label{sec:iclmeth}

The intra-cluster light (ICL) represents an important component of the cluster mass distribution. In addition, it is a unique tracer of a system's dynamical history and its underlying dark-matter distribution, as demonstrated in recent works (e.g., \citealt{Montes2014,Montes2019,Montes2022,Montes2022b,Gonzalez2021,Deason2021}). While the ICL has so far proven extremely difficult to detect and study with ground- and space-based telescopes, the exceptional sensitivity of \textit{JWST}'s detectors holds great promise for the detection of these extended, yet extremely low-surface-brightness features.

In order to enhance faint, diffuse emission, we apply a running median filtering with a 21$\times$21 pixel box size. The resulting image is shown in Fig.~\ref{fig:icl}.

\subsection{Strong-lensing mass modeling}
\label{Methodology}

We derive a mass model for SMACS\,J0723 based on strong-lensing constraints identified in the cluster core, using the publicly available mass modeling algorithm \lenstool\ \citep{Jullo2007}. We provide a short summary of our approach here and refer the reader to \cite{Kneib1996}, \cite{Smith2005}, \cite{Verdugo2011} and \cite{Richard2011} for more details. The cluster mass distribution is modeled as a series of parametric dual pseudo-isothermal ellipsoidal (dPIE, \citealt{Eliasdottir2007}) dark matter halos with seven free parameters: the position $\Delta\alpha$, $\Delta\delta$ relative to a reference location; ellipticity $.pdfilon$; position angle $\theta$; normalization $\sigma_{0,lt}$; truncation radius $r_{cut}$; and core radius $r_{core}$. We use as input constraints the positions of prominent light peaks in each lensed image, as well as their spectroscopic redshifts where available (see Section~\ref{sec:cstr}) and large flat priors otherwise. The \lenstool\ algorithm uses a Monte Carlo Markov Chain (MCMC) formalism to explore the available parameter space and identifies the best fit as the set of parameter values that minimizes the scatter between the observed and predicted image-plane positions of the identified lensed features. 

The lens plane is modeled as a combination of cluster-scale and galaxy-scale dPIE halos. For the cluster-scale DM halos, \review{we fix the truncation radius ($r_{cut}$) at 1500\,kpc. This radius typically lies outside the strong-lensing region and therefore cannot be well-constrained using our model. We refer to \cite{Chang2018} and reference therein for relevant insights on choosing this radius as the truncation radius.} All other parameters are optimized unless indicated otherwise. 

Galaxy-scale halos represent the contribution to the lensing potential from cluster member galaxies \citep[e.g.,][]{NatarajanKneib1997,2019MNRAS.483.3082J,Sharon2020}. Their positional parameters ($\Delta\alpha$, $\Delta\delta$; $.pdfilon$; $\theta$) are fixed at their observed values as measured with Source Extractor \citep{SEx}, (note that this subset includes ellipticity and position angle). The cluster-member catalog relies on \HST \ photometry, as the two filters F606W and F814W (which straddle the Balmer break at the cluster redshift) provide a color gradient that allows us to isolate cluster member galaxies that form the so-called red sequence \citep{Gladders2000} shown in Fig.~\ref{fig:zhist}. \review{We identify 144 galaxies.} We also independently identify 26 cluster member galaxies from MUSE spectroscopy \review{ranging from z$=$0.3727 to z$=$0.3981 based on the clear overdensities in redshift as shown in the histogram Fig.~\ref{fig:zhist}} and note that four of these fall outside the color range chosen for our red-sequence selection \review{and where included in our cluster member catalogue.}

To keep the number of optimized model parameters manageable in terms of computing time, we do not model the parameters of the galaxy-scale potentials individually but scale them to their observed $i$-band luminosity (using the Source Extractor output MAG\_AUTO value) with respect to L$^*$ (mag$_{\rm F814W}$=19.12), using a parameterized mass-luminosity scaling relation with a constant mass-luminosity ratio (see \citealt{NatarajanKneib1997,Limousin2007} and discussions therein on the validity of this approach), leaving only the cut radius, $r_{cut}$, and the fiducial central velocity dispersion, $\sigma_{0,lt}$, free to vary. We note that L$^*$ is degenerate with the $\sigma_{0,lt}$ normalization and offers flexibility. The BCG is modeled separately, since extremely luminous central cluster galaxies often do not follow the aforementioned general scaling relation (\citealt{Newman1,Newman2}). In addition, we separately model the cluster member galaxy at (R.A.$=110.8402908$ Decl.$=-73.4559518$) which, being closest to the lensed image  dubbed "The Sparkler" (image 2.2), has a disproportionate influence on the lens model (see \citealt{Claeyssens2022,Mowla2022} and references therein for a more detailed discussion of the Sparkler). Altogether we thus jointly optimize 146 galaxies using our constant mass-luminosity relation.

%The models we construct and present here are publicly available\footnote{\url{https://www.dropbox.com/sh/3iatmz5k4hafzqf/AAAh0JvLgpBVoLp6qsxYZkFGa?dl=0}}; the linked-to website will be constantly updated and point to other additional repositories as they become available\footnote{\url{https://sites.google.com/view/guillaume-mahler-astronomer/smacs0723-precise-lens-modelling}}.

The models we construct and present here are publicly available\footnote{\url{https://github.com/guillaumemahler/SMACS0723-mahler2022}}; the linked-to website will be constantly updated.

\subsection{X-ray analysis}
\label{sec:xrays}

To recover the properties of the gaseous intracluster medium (ICM) from the existing short \textit{Chandra} X-ray observation of SMACS\,J0723, we model the spectrum of the emission with the Astrophysical Plasma Emission Code (APEC)\footnote{\url{http://atomdb.org/}}, adopting abundance ratios as provided by \citet{2009ARA&A..47..481A}. To account for foreground absorption, we complement this main spectral component with a photoelectric-absorption model\footnote{\url{https://heasarc.gsfc.nasa.gov/docs/xanadu/xspec/manual/XSmodelPhabs.html}}. The contribution from background emission is incorporated by creating an empirical model of the blank-sky background with B-spline functions whose coefficients were obtained through a fit of the blank-sky spectrum for the ACIS-I CCD on which the cluster is observed. We then keep the shape of the background spectrum constant in the fitting procedure and allow only its normalization to vary.

We perform all modeling within the \textsc{Sherpa} fitting environment \citep{sherpa} combined with the \textsc{Python} wrapper of the \textsc{MultiNest} nested sampling package \citep{pymultinest,multinest} to explore the parameter space of our model in the $0.5$-$8.0$ keV energy band. As appropriate for the mostly low photon statistics per bin, we use a Poisson likelihood similar to \textsc{cstat}\footnote{\url{https://heasarc.gsfc.nasa.gov/xanadu/xspec/manual/XSappendixStatistics.html}}. Depending on the fit, not all emission model parameters are left free to vary. We consider the background normalization a nuisance parameter and marginalize over it in our best-fit estimates for all physical model parameters.

\section{Results}
\label{sec:results}

\subsection{Cluster galaxies}
\label{sec:clustergal}

In order to obtain an independent assessment of the dynamical state of SMACS\,J0723 as probed by the spatial and velocity distribution of the system's member galaxies, we examine the MUSE data cube and extract a catalog of 26 spectroscopically confirmed cluster members (Table~\ref{tab:cmembers}).

\begin{figure}
    \centering
    \includegraphics[width=\columnwidth]{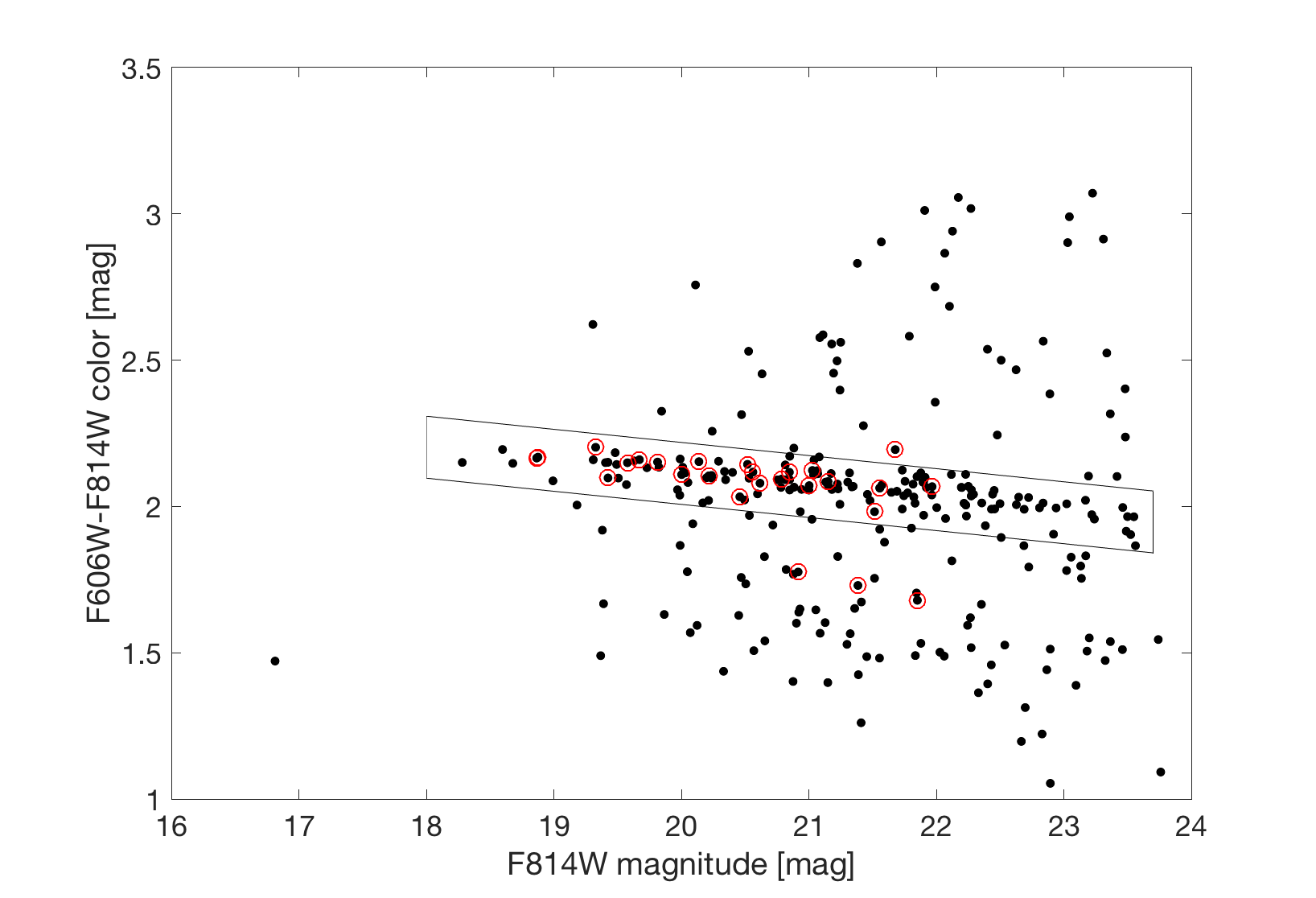}
    \includegraphics[width=\columnwidth]{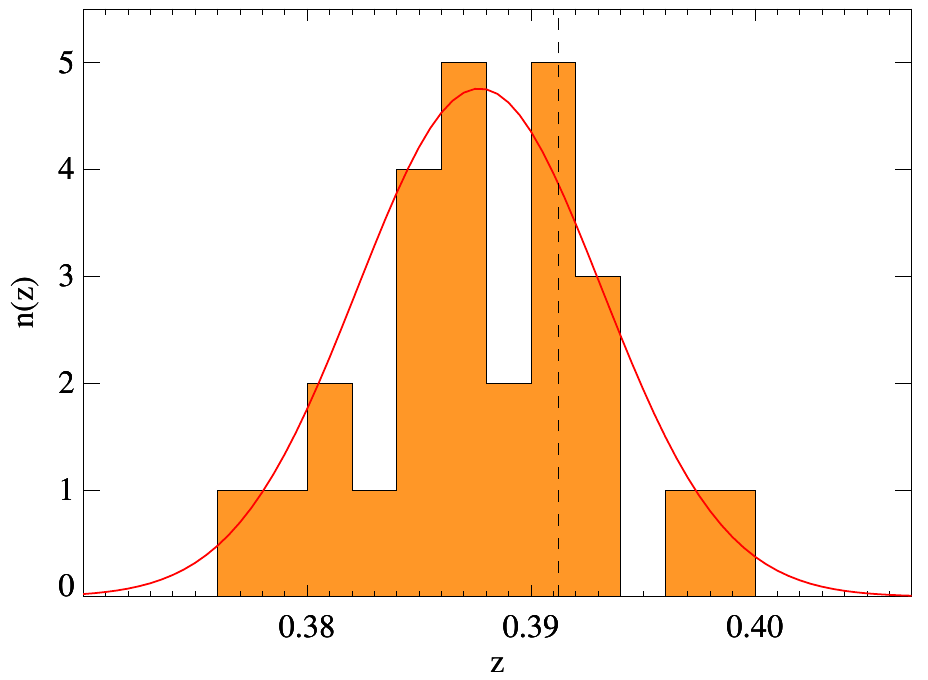}
    \caption{  Top: Color-magnitude diagram of galaxies in the field of view of SMACS J0723. The red sequence of cluster member galaxies is clearly visible since the two filters used, F606W and F814W, straddle the Balmer break of massive elliptical galaxies at that redshift. The 26 spectroscopically confirmed cluster members from MUSE spectroscopy (open red circles) are overplotted. The rectangular shape shows the selection of the \review{144} cluster member galaxies used in our lens model. Additionally, the four spectroscopically identified cluster members are included in our final cluster member catalog. 
    Bottom: Histogram of the redshifts of the 26 cluster members identified spectroscopically within the MUSE data cube \review{ranging from 0.3727 to 0.3981}. Overlaid is the best Gaussian model which determines the cluster velocity dispersion. The vertical dashed line marks the location of the BCG in redshift space, $z_{BCG}=0.3912$, which is displaced from the systemic redshift of the cluster, $z=0.3877$ \review{corresponding to the mean redshift} (see~\autoref{sec:clustergal}
    }
    \label{fig:zhist}
\end{figure}

Using the \textsc{ROSTAT} package of \citet[][]{Beers1990} we derive an improved cluster redshift of $z=0.3877$ for SMACS\,J0723 and determine the cluster velocity dispersion as $\sigma=1180_{-180}^{+160}$ km s$^{-1}$. We show the corresponding redshift histogram in Fig.~\ref{fig:zhist}. Within the statistical uncertainties set by the small sample size, the radial velocity distribution exhibits no significant substructure indicative of an active merger along an axis that lies close to our line of sight. We note, however, that the radial velocity of the BCG is clearly offset from the centroid of the distribution; the implied peculiar velocity might reflect incomplete relaxation after a potentially recent line-of-sight merger.

\begin{table}
    \centering
    \begin{tabular}{ccc}
        R.A. (deg) & Decl. (deg) & $z$ \\ \hline
   110.80001 &   -73.45269 &  0.3791 \\
   110.80062 &   -73.44852 &  0.3936 \\
   110.80247 &   -73.45867 &  0.3904 \\
   110.80451 &   -73.45615 &  0.3862 \\
   110.81613 &   -73.45119 &  0.3841 \\
   110.81726 &   -73.44940 &  0.3930 \\
   110.81824 &   -73.45462 &  0.3908 \\
   110.81841 &   -73.44827 &  0.3936 \\
   110.81852 &   -73.45524 &  0.3848 \\
   110.82437 &   -73.45991 &  0.3809 \\
   110.82514 &   -73.45454 &  0.3767 \\
   110.82571 &   -73.45869 &  0.3885 \\
   110.82639 &   -73.45499 &  0.3909 \\
   110.82688 &   -73.45463 &  \hspace*{9.2mm}0.3912 (BCG) \\
   110.83269 &   -73.45691 &  0.3867 \\
   110.83683 &   -73.45652 &  0.3981 \\
   110.83763 &   -73.45617 &  0.3895 \\
   110.83780 &   -73.45360 &  0.3864 \\
   110.84009 &   -73.45587 &  0.3908 \\
   110.84564 &   -73.45134 &  0.3845 \\
   110.84875 &   -73.46031 &  0.3970 \\
   110.85310 &   -73.45666 &  0.3838 \\
   110.85378 &   -73.45006 &  0.3844 \\
   110.85506 &   -73.45020 &  0.3864 \\
   110.85574 &   -73.45574 &  0.3815 \\
   110.85626 &   -73.45070 &  0.3872 \\
    \end{tabular}
    \caption{Right ascension and declination (J2000) as well as redshifts of the 26 cluster members identified in the MUSE observation of the core of SMACS\,J0723.}
    \label{tab:cmembers}
\end{table}

\subsection{ICL}
\label{sec:iclres}

We examine the filtered NIRCam image of SMACS\,J0723 shown in Fig.~\ref{fig:icl} in search of unusual low-surface-brightness features and note diffuse excess emission west of the cluster core but also past the far eastern extension of the ICL halo of the BCG. Although the physical nature and origin of such excess ICL are not immediately clear, we mark these areas as locations of potential minor mass concentrations within the cluster lens that are not associated with either over-densities of cluster members or excess X-ray emission and are thus not readily identifiable by other means.

\subsection{Strong-lensing constraints}
\label{sec:cstr}

\begin{figure*}
    \centering
    \includegraphics[width=\textwidth]{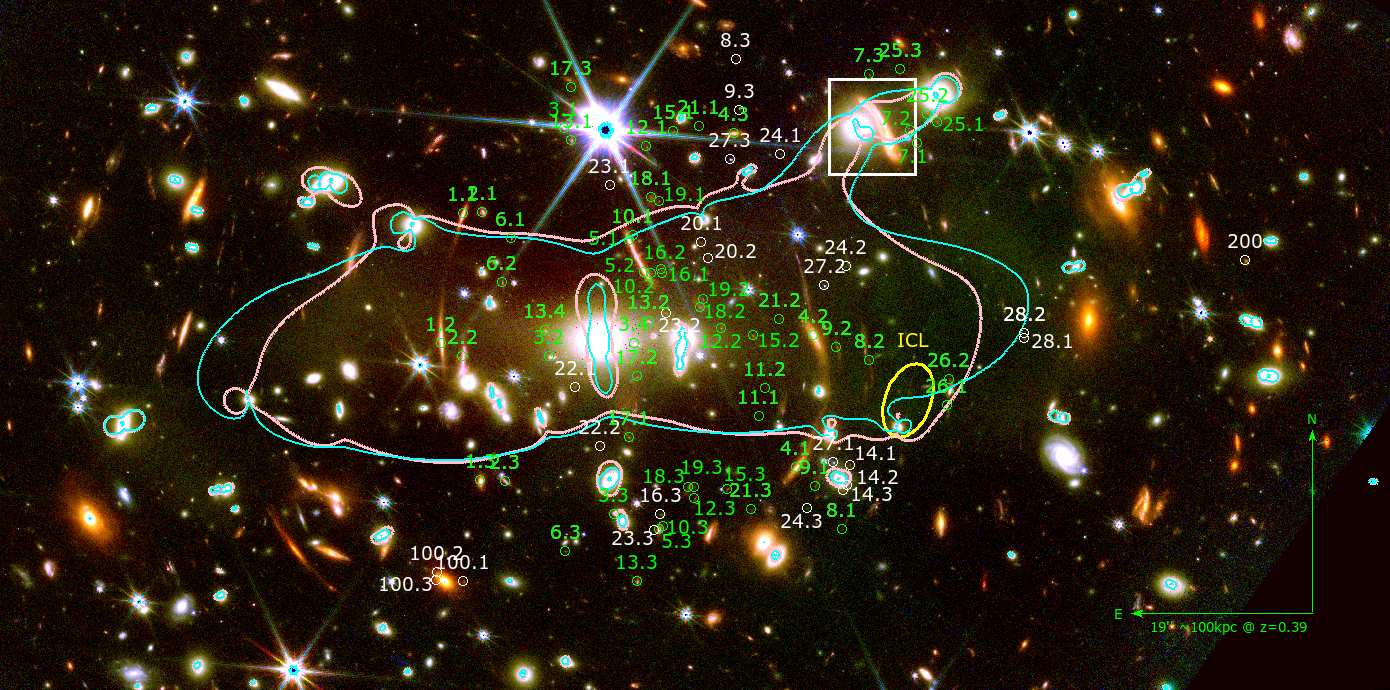}
   \includegraphics[width=\textwidth]{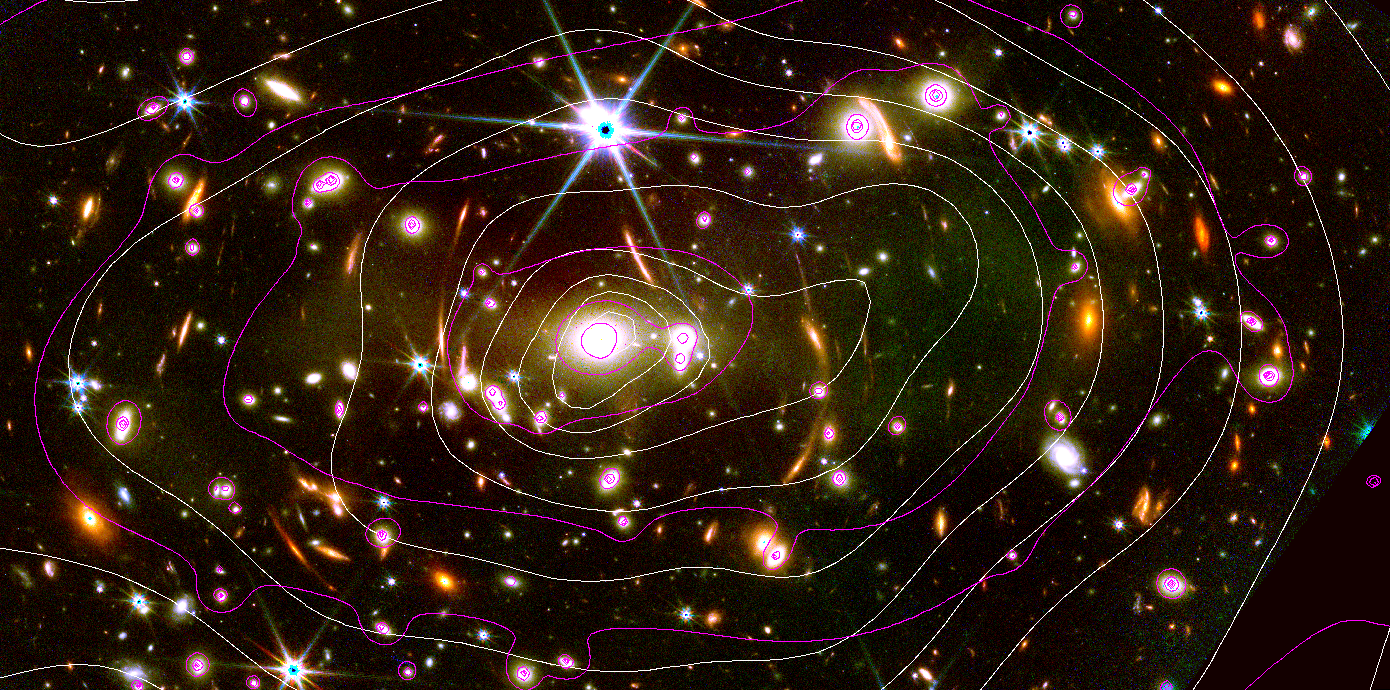}
    \caption{Top: Color image of SMACS\,J0723 with multiple-image systems used in our models marked  by green circles and all other candidates marked as white circles. Also shown are the critical curves for a source at redshift $z=9$; in cyan for the single-component lens model and in pink for our final model that includes one additional mass clump marked by excess ICL at the location of the yellow ellipse. The white square highlights the 'Beret' galaxy, a highly stretched spiral at $z=1.16$ that is only partly multiply imaged (Sect.\,\ref{sec:cstr}). Bottom: Color image of SMACS\,J0723 with mass contours (in magenta) and X-ray surface-brightness contours (in white) overlaid. We note the visual similarity in ellipticity and asymmetrical distribution along the East-West axis. }
    \label{fig:model}
\end{figure*}

The strong-lensing constraints for our lens models are given by the image-plane locations of multiple images of lensed sources, identified either in previous \textit{HST} images or in the new \textit{JWST} observations. 

\citet{Golubchik2022} identify five arc candidates in the field of SMACS\,J0723 and report three multiple-image systems that have spectroscopic redshifts. We confirm all of these in our examination of all available data and identify 16 additional multiple-image systems. Through careful inspection of the MUSE datacube (Section~\ref{sec:Spectroscopy}), we also secure an additional spectroscopic redshift for one of the systems photometrically identified by \cite{Golubchik2022}; System 3 (at $z = 1.9914$).

\begin{figure}
\begin{center}
	\includegraphics[width=\columnwidth]{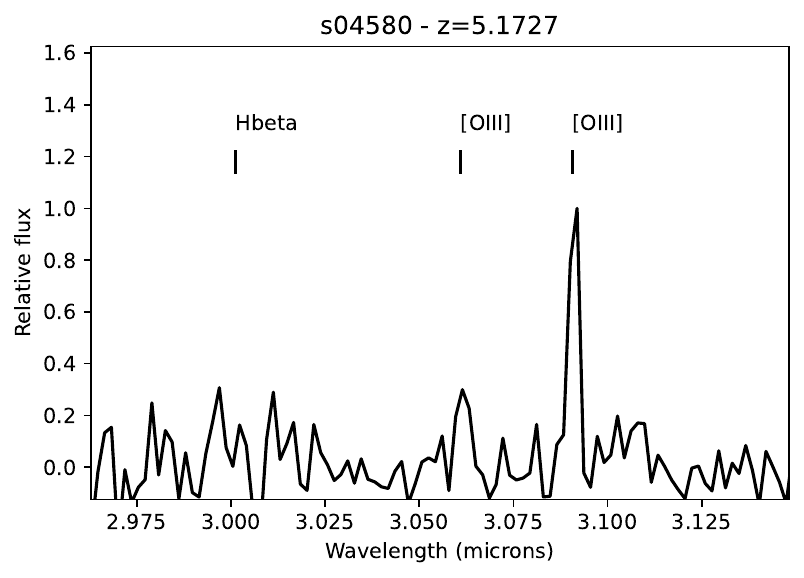}
	\includegraphics[width=\columnwidth]{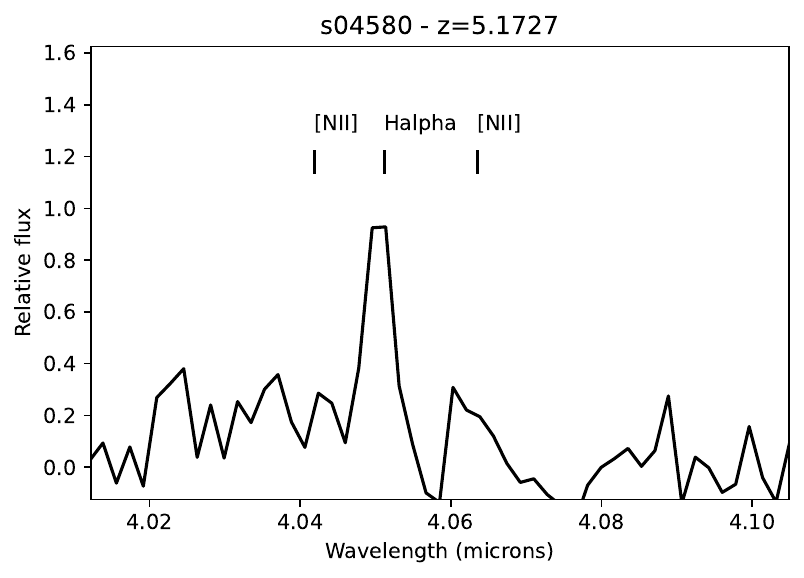}
    \caption{Identified emission lines in the NIRSpec/G395m spectrum of image 7.1. The detection of strong [OIII]5007\AA\ and H$\alpha$ lines, accompanied by weaker [OIII]4959\AA\ and [NII] emission, makes this redshift determination robust.}
    \label{spec:sys17}
\end{center}
\end{figure}

Initial inspection of the NIRSpec MSA spectroscopic data yields an additional spectroscopic constraint by confirming a star-forming region in image 7.1 to be at $z=5.1727$, the highest redshift of any spectroscopically confirmed multiple image in this cluster (Fig.~\ref{spec:sys17}). We also confirm the redshift for the 'Beret' galaxy (Fig. \ref{fig:model}), a highly stretched spiral galaxy that is only partially multiply imaged, as $z=1.16$; however, we do not include this image as a modeling constraint.

All individual images are marked in Fig.~\ref{fig:model}, and Table~\ref{tab:cstr} summarizes the positions and spectroscopic redshifts where available. Although the identification of systems without spectroscopic confirmation for all individual images should in principle be considered tentative, we propose to adopt Systems 1, 2 and 3, as secure identifications, in view of their unique morphology, which is identical for all of their multiple images.

\startlongtable
\begin{deluxetable}{ccccc} 
\tablecolumns{5} 
\tablecaption{Securely identified multiple-image systems, denoted by a "System.ID" nomenclature. "System" specifies the group of images originating from the same source galaxy, whereas "ID" refers to the name of the individual image. R.A. and Decl. are the right ascension and declination (J2000) of the image. $z$ is the measured spectroscopic redshift. \review{Redshifts with error bars denote the median model-optimized redshift and the 68\% confidence interval. Systems with $\dagger$ symbols are not used as constraints in this model.}
$\mu$ is the magnification at the location of the observed constraints. 
Where errors are listed for $\mu$, the cited values are the \review{median} magnification and the %1$\sigma$ statistical 
68\% confidence interval from the lens-model optimization.}
	\label{tab:cstr}
\tablehead{\colhead{Sys.} &
            \colhead{R.A. [deg]}    & 
            \colhead{Decl. [deg]}    & 
            \colhead{z}     & 
            \colhead{$\mu$ } \\ [-8pt]
%            \\        
            \colhead{ID} &
            \colhead{J2000}     & 
            \colhead{J2000}    & 
            \colhead{}       & 
            \colhead{}             }
\startdata 
\hline
1.1 & 110.8407240 & -73.4510787 & 1.449 &  $5.5^{+0.7}_{-0.6}$ \\ 
1.2 & 110.8429489 & -73.4548399 & 1.449 &  $11.4^{+2.1}_{-2.0}$ \\ 
1.3 & 110.8389887 & -73.4587844 & 1.449 &  $5.2^{+0.6}_{-0.5}$ \\ 
\hline
2.1 & 110.8387288 & -73.4510508 & 1.3779 &  $5.1^{+0.6}_{-0.5}$ \\ 
2.2 & 110.8407771 & -73.4552122 & 1.3779 &  $11.3^{+2.1}_{-1.9}$ \\ 
2.3 & 110.8364983 & -73.4588136 & 1.3779 &  $4.8^{+0.5}_{-0.5}$ \\ 
\hline
3.1 & 110.8305036 & -73.4486312 & 1.9914 &  $3.0^{+0.3}_{-0.2}$ \\ 
3.2 & 110.8319988 & -73.4552022 & 1.9914 &  $3.0^{+0.4}_{-0.3}$ \\ 
3.3 & 110.8254393 & -73.4597767 & 1.9914 &  $7.8^{+1.5}_{-0.8}$ \\ 
3.4 & 110.8233893 & -73.4548350 & 1.9914 &  $1.9^{+0.3}_{-0.3}$\\ 
\hline
4.1 & 110.8069982 & -73.4584308 & $2.31_{-0.10}^{0.12}$ & $6.9^{+0.6}_{-0.6}$ \\ 
4.2 & 110.8052367 & -73.4546325 & \nodata & $14.0^{+2.5}_{-1.9}$\\ 
4.3 & 110.8132881 & -73.4487869 & \nodata & $4.4^{+0.4}_{-0.4}$ \\ 
\hline
5.1 & 110.8238908 & -73.4518820 & 1.425 & $18.3^{+2.9}_{-2.4}$ \\ 
5.2 & 110.8223529 & -73.4527831 & 1.425 & $20.0^{+3.0}_{-2.2}$ \\ 
5.3 & 110.8209254 & -73.4602058 & 1.425 & $3.0^{+0.2}_{-0.2}$ \\ 
\hline
6.1 & 110.8358540 & -73.4518199 & $1.70_{-0.03}^{+0.04}$ & $14.2^{+2.8}_{-2.1}$ \\ 
6.2 & 110.8367611 & -73.4530868 & \nodata & $12.7^{+1.5}_{-1.0}$ \\ 
6.3 & 110.8303933 & -73.4608436 & \nodata & $3.0^{+0.2}_{-0.2}$ \\ 
\hline
7.1 & 110.7947604 & -73.4490975 & 5.17 & $20.3^{+9.9}_{-4.1}$ \\ 
7.2 & 110.7954442 & -73.4487211 & 5.17 & $23.3^{+21.8}_{-6.7}$ \\ 
7.3 & 110.7996039 & -73.4470866 & 5.17 & $5.4^{+1.0}_{-0.5}$ \\ 
\hline
8.1 & 110.8023784 & -73.4602055 & $14.39_{-2.11}^{+1.17}$ & $4.7^{+0.9}_{-0.5}$ \\ 
%14.3895 11.7580 15.9331 +-1.5760 +0.0474 -2.4424
8.2 & 110.7995598 & -73.4553501 & \nodata & $9.6^{+1.3}_{-1.8}$ \\ 
$^\dagger$8.3 & 110.8130564 & -73.4466651 & \nodata & \nodata\\ 
\hline
9.1 & 110.8050637 & -73.4589656 & $3.01_{-0.21}^{+0.25}$ & $7.2^{+0.8}_{-0.8}$ \\ 
9.2 & 110.8028896 & -73.4549564 & \nodata & $16.0^{+4.5}_{-2.3}$\\ 
$^\dagger$9.3 & 110.8127004 & -73.448125 & \nodata &\nodata \\ 
\hline
10.1 & 110.8235289 & -73.4517392 & $1.43_{-0.02}^{+0.02}$ & $15.2^{+2.5}_{-1.9}$ \\ 
10.2 & 110.8216192 & -73.4528243 & \nodata & $16.6^{+2.3}_{-1.9}$ \\ 
10.3 & 110.8205119 & -73.4601152 & \nodata & $3.0^{+0.2}_{-0.2}$ \\ 
\hline
11.1 & 110.8107306 & -73.4569574 & $1.73_{-0.09}^{+0.11}$ & $23.5^{+7.3}_{-3.1}$ \\ 
11.2 & 110.8101464 & -73.4561599 & \nodata & $22.9^{+7.3}_{-3.8}$ \\ 
\hline
12.1 & 110.8221364 & -73.4491504 & $1.81_{-0.06}^{+0.07}$ & $3.8^{+0.4}_{-0.3}$ \\ 
12.2 & 110.8146179 & -73.4544119 & \nodata & $3.6^{+0.5}_{-0.5}$ \\ 
12.3 & 110.8173093 & -73.459317 & \nodata & $4.0^{+0.4}_{-0.3}$\\ 
\hline
13.1 & 110.8297224 & -73.4489907 & $3.34_{-0.3}^{+0.49}$ & $3.9^{+0.4}_{-0.3}$ \\ 
13.2 & 110.821915 & -73.4542067 & \nodata & $3.3^{+0.5}_{-0.4}$ \\ 
13.3 & 110.823115 & -73.46170.5 & \nodata & $3.0^{+0.2}_{-0.2}$ \\ 
13.4 & 110.8324286 & -73.4544642 & \nodata & $3.0^{+0.5}_{-0.4}$ \\ 
\hline
$^\dagger$14.1 & 110.8015568 & -73.4583546 & \nodata & \nodata\\ 
$^\dagger$14.2 & 110.8018148 & -73.458948 & \nodata & \nodata\\ 
$^\dagger$14.3 & 110.802227 & -73.4590843 & \nodata &\nodata \\ 
\hline
15.1 & 110.8193895 & -73.4487436 & $2.04_{-0.08}^{+0.09}$ & $4.3^{+0.4}_{-0.4}$\\ 
15.2 & 110.8113813 & -73.4546235 & \nodata & $5.1^{+0.6}_{-0.6}$\\ 
15.3 & 110.8139705 & -73.4590522 & \nodata & $4.6^{+0.4}_{-0.4}$\\ 
\hline
16.1 & 110.82062 & -73.4527181 & $1.09_{-0.03}^{+0.03}$ & $214.3^{+262.7}_{-23.5}$\\ 
16.2 & 110.820525 & -73.4528156 & \nodata & $205.1^{+235.3}_{-20.2}$\\ 
$^\dagger$16.3 & 110.8207626 & -73.4597746 & \nodata &\nodata \\ 
\hline
17.1 & 110.8239479 & -73.4575528 & $2.12_{-0.09}^{+0.11}$ & $15.3^{+2.7}_{-2.3}$\\ 
17.2 & 110.8231354 & -73.4558083 & \nodata & $7.9^{+0.8}_{-0.8}$ \\ 
17.3 & 110.8297769 & -73.4474619 & \nodata & $2.5^{+0.2}_{-0.2}$ \\ 
\hline
18.1 & 110.8216711 & -73.4506362 & $1.37_{-0.03}^{+0.03}$ & $5.7^{+0.7}_{-0.5}$ \\ 
18.2 & 110.816745 & -73.4537968 & \nodata & $6.8^{+0.9}_{-0.7}$ \\ 
18.3 & 110.817934 & -73.4590101 & \nodata & $3.7^{+0.3}_{-0.3}$ \\ 
\hline
19.1 & 110.8208804 & -73.4507461 &  $1.37_{-0.03}^{+0.03}$ & $6.3^{+0.3}_{-0.3}$ \\ 
19.2 & 110.8164058 & -73.4535733 & \nodata & $7.7^{+1.1}_{-0.8}$\\ 
19.3 & 110.8173046 & -73.4589942 & \nodata & $3.7^{+0.3}_{-0.3}$\\ 
\hline
$^\dagger$20.1 & 110.8165814 & -73.4519445 & \nodata & \nodata\\ 
$^\dagger$20.2 & 110.8159392 & -73.4523932 & \nodata & \nodata\\ 
\hline
21.1 & 110.8168354 & -73.448577 & $2.60_{-0.14}^{+0.17}$ & $4.1^{+0.4}_{-0.4}$ \\ 
21.2 & 110.8086654 & -73.4541442 & \nodata & $6.1^{+0.7}_{-0.7}$\\ 
21.3 & 110.8115827 & -73.4596446 & \nodata & $4.2^{+0.3}_{-0.3}$\\ 
\hline
$^\dagger$22.1 & 110.82934 & -73.4561204 & \nodata & \nodata\\ 
$^\dagger$22.2 & 110.826863 & -73.4578161 & \nodata & \nodata\\ 
\hline
$^\dagger$23.1 & 110.8258363 & -73.4502839 & \nodata & \nodata\\ 
$^\dagger$23.2 & 110.8201612 & -73.4539789 & \nodata & \nodata\\ 
$^\dagger$23.3 & 110.8213975 & -73.4602314 & \nodata & \nodata\\ 
\hline
$^\dagger$24.1 & 110.8085708 & -73.4494083 & \nodata & \nodata\\ 
$^\dagger$24.2 & 110.8019579 & -73.4526322 & \nodata & \nodata\\ 
$^\dagger$24.3 & 110.8058921 & -73.4595997 & \nodata & \nodata\\ 
\hline
25.1 & 110.7927038 & -73.4484814 & $3.93_{-1.01}^{+1.65}$ & $13.7^{+4.7}_{-2.5}$\\ 
25.2 & 110.7936842 & -73.4482439 & \nodata & $10.6^{+4.0}_{-2.1}$ \\ 
25.3 & 110.7964129 & -73.4469406 & \nodata & $5.1^{+0.8}_{-0.6}$ \\ 
\hline
26.1 & 110.7917089 & -73.4566332 & $2.88_{-1.15}^{+1.35}$ & $60.6^{+49.9}_{-31.7}$ \\ 
26.2 & 110.7914913 & -73.4558973 & \nodata & $64.8^{+77.8}_{-5.5}$ \\ 
\hline
$^\dagger$27.1 & 110.8032246 & -73.4582886 & \nodata & \nodata\\ 
$^\dagger$27.2 & 110.8041292 & -73.4531883 & \nodata & \nodata\\ 
$^\dagger$27.3 & 110.8136692 & -73.4495378 & \nodata & \nodata\\ 
\hline
$^\dagger$28.1 & 110.7839071 & -73.4547219 & \nodata & \nodata \\ 
$^\dagger$28.2 & 110.7838671 & -73.4545531 & \nodata & \nodata \\ 
\hline
$^\dagger$100.1 & 110.840764 & -73.46169 & \nodata & \nodata\\ 
$^\dagger$100.2 & 110.8433794 & -73.4614539 & \nodata & \nodata\\ 
$^\dagger$100.3 & 110.843516 & -73.4616658 & \nodata & \nodata\\ 
\hline
$^\dagger$200 & 110.7615033 & -73.4524747 & \nodata & \nodata\\ 
\hline
\enddata
\end{deluxetable} 
%The value in the table are updated with the latest model S5_moreMCMC
\subsection{Mass distribution}

\begin{center}
\begin{table*}
\caption{Candidate Lens Models and Output Parameters}
\label{tab:model_param}
\begin{tabular}{lrcccccccc}
\hline
Model name & Component & $\Delta\alpha^{\rm ~a}$ & $\Delta\delta^{\rm ~a}$ & $\varepsilon^{\rm ~b}$ & $\theta$ & $\sigma_{0,lt}^{\rm ~c}$ & r$_{\rm cut}$ & r$_{\rm core}$\\
(Fit statistics) &  & (\arcsec) & (\arcsec) &   & ($\deg$) & (km\ s$^{-1}$) & (kpc) & (kpc)\\ 
\hline 
\review{Fiducial model} & Cluster halo & 2.82$^{+0.92}_{-0.9}$ & 1.31$^{+0.25}_{-0.22}$ & 0.67$^{+0.05}_{-0.05}$ & 8.1$^{+0.77}_{-0.71}$ & 983.32$^{+31.85}_{-35.39}$ & [1500.0] & 17.96$^{+1.45}_{-1.44}$\\ 
rms = 0.32\arcsec\  k = 46 & BCG & [0.0] & [0.0] & -- & [29.2] & 292.09$^{+23.36}_{-19.75}$ & 56.39$^{+25.86}_{-32.54}$ & 0.44$^{+0.42}_{-0.28}$\\ 
 $\chi^2/\nu$ = 1.0 dof = 32 & $^{\rm d}$NW clump & -30.8$^{+1.63}_{-1.8}$ & 22.28$^{+1.48}_{-0.79}$ & 0.35$^{+0.29}_{-0.24}$ & 6.53$^{+58.6}_{-70.74}$ & 154.59$^{+24.27}_{-29.65}$ & 34.77$^{+13.53}_{-17.14}$ & 0.92$^{+0.56}_{-0.5}$\\ 
$\log$($\mathcal{L}$) = -28 & ICL clump & -34.51$^{+2.29}_{-5.52}$ & -5.4$^{+1.97}_{-1.12}$ & 0.5$^{+0.23}_{-0.21}$ & 40.9$^{+20.81}_{-16.81}$ & 375.86$^{+51.21}_{-55.89}$ & 158.97$^{+22.02}_{-45.46}$ & 7.07$^{+1.83}_{-3.71}$\\ 
 $\log$($\mathcal{E}$) = -133 & $^{\rm e}$CM gal & [13.64] & [-4.42] & 0.31$^{+0.2}_{-0.21}$ & -0.42$^{+62.04}_{-62.99}$ & 26.91$^{+32.69}_{-18.62}$ & 19.77$^{+12.51}_{-12.53}$ & 1.02$^{+0.59}_{-0.66}$\\ 
BIC = 256 AICc = 287 & $L^{*}$ Galaxy & -- & -- & -- & -- & 144.8$^{+13.1}_{-11.9}$ & 67.5$^{+20.6}_{-18.5}$ & [0.15]\\ 
\hline 
\review{Comparison model} & Cluster Halo & -4.44$^{+2.12}_{-2.05}$ & 1.04$^{+0.4}_{-0.41}$ & 0.86$^{+0.03}_{-0.04}$ & 183.86$^{+0.68}_{-0.71}$ & 1079.89$^{+44.21}_{-37.73}$ & [1500.0] & 21.69$^{+3.34}_{-2.91}$\\ 
rms = 0.85\arcsec\  k = 39 & BCG & [0.0] & [0.0] & -- & [29.2] & 362.25$^{+27.57}_{-25.55}$ & 12.54$^{+3.1}_{-3.39}$ & 0.28$^{+0.2}_{-0.2}$\\ 
 $\chi^2/\nu$ = 1.0 dof = 39 & $^{\rm d}$NW clump & -28.35$^{+1.81}_{-2.26}$ & 22.38$^{+2.14}_{-1.4}$ & 0.28$^{+0.23}_{-0.15}$ & -5.79$^{+61.23}_{-56.0}$ & 268.71$^{+20.96}_{-29.54}$ & 57.71$^{+22.77}_{-19.9}$ & 4.06$^{+5.56}_{-2.77}$\\ 
$\log$($\mathcal{L}$) = -142 & $^{\rm e}$CM gal  & [13.64] & [-4.42] & 0.54$^{+0.31}_{-0.36}$ & 4.21$^{+59.91}_{-66.9}$ & 91.83$^{+40.79}_{-51.69}$ & 19.68$^{+12.22}_{-13.28}$ & 0.58$^{+0.25}_{-0.26}$\\ 
 $\log$($\mathcal{E}$) = -202 & $L^{*}$ Galaxy & -- & -- & -- & -- & 169.6$^{+38.1}_{-36.3}$ & 48.7$^{+17.9}_{-23.3}$ & [0.15]\\ 
BIC = 454 AICc = 444 & -- & -- & -- & -- & -- & -- & -- & --\\ 
\hline 
\end{tabular}
%\medskip\\
$^{\rm a}$ $\Delta\alpha$ and $\Delta\delta$ are the relative position to the reference coordinate point: ($\alpha$ = 110.82675, $\delta$ = -73.454628)\\%~~~~~~~~~~~~~~~~~~~~~~~~~~~~~~~\\[1pt]
$^{\rm b}$ Ellipticity ($\varepsilon$) is defined to be $(a^2-b^2) / (a^2+b^2)$, where $a$ and $b$ are the semi-major and semi-minor axes of the ellipse\\%~~~~~~\\[1pt]
$^{\rm c}$ $\sigma_{0,lt}$ is the normalization parameter and represents a fiducial central velocity dispersion as defined in the dPIE parametrization within lenstool\\%\\[1pt]
$^{\rm d}$ NW clump refers to the additional north-western clump near system 7, 25 and the galaxy nicknamed "the Beret"\\%\\[1pt]
$^{\rm e}$ ``CM gal." refers to the galaxy near system 2.2 (the Sparkler)\\%\\[1pt]
$^{\rm ~f}$ k is the number of free parameter in the model \\%\\[1pt]
Quantities in brackets are fixed parameters. \review{Other output quantities are the median value and the 68\% confidence interval from the model optimization}~~~~~~~~~~~~~~~~~~~~~~~~~~~~~~~~~~~~~~~~~~~~~~~~~~~~~~~~~~~~~~~~~~~~~~~~~~~~~~~~~~~~~~~~~~~~~~~~~~~~~~~~~
\end{table*}
\end{center}

\subsubsection{Excess mass}
\label{sec:diffmass}

The presence of two bright galaxies north-west of the BCG motivates the inclusion of an additional large-scale halo in our model to better accommodate two nearby multiply imaged galaxies (Systems 7 and 25, See Section \ref{sec:cstr}). Moreover, while we see no significant substructure in the distribution of cluster galaxies west and south-west of the BCG, we observe an extension of the ICL in these directions. The presence of this excess diffuse light (highlighted in Fig.~\ref{fig:icl} and discussed in Sections~\ref{sec:iclmeth} and \ref{sec:iclres}) causes us to add a second large-scale mass component which proves crucial to reproducing the observed lensing geometry of Systems 8 and 26. Fig.\,\ref{fig:model} shows the location of the additional component, referred to as the ``ICL clump" in Table\,\ref{tab:model_param}.

To assess the importance of this additional component to our mass model, we run two models with parameters as listed in Table \ref{tab:model_param}: one with only a cluster-scale halo around the BCG (Comparison Model in Table\,\ref{tab:model_param}), and another one including the additional large-scale halos described above (Fiducial Model in Table~\ref{tab:model_param}). Proceeding in our analysis, as described below, we only use the most complex model since it provides a better overall RMS and Bayesian Information Criterion (BIC; \citealt{Schwarz1978}), both criteria used in previous works (e.g. \citealt{Acebron2017,Collett2017,Lam2018})

\subsubsection{Comparison with other mass models}

We compare the results of our improved strong-lensing analysis with models from previous works on SMACS\,J0723. Two of these are from the public release of RELICS cluster models \citep{Coe2019}, derived using the GLAFIC lens mapping package and a \lenstool\ model detailed in \citealt{Sharon2022}, respectively. In addition, we compare our results with those from the recent analysis by \citet{Golubchik2022}, performed using the Light Traces Mass (LTM) software. 

Table\,\ref{tab:mass} lists and compares the masses from all existing lens models for SMACS\,J0723 at three different radii: 128\,kpc, 200\,kpc, and 400\,kpc. Here, 128\,kpc corresponds to the largest cluster-centric distance of the strong-lensing constraints commonly used by all models (this multiple-image system is labeled System 4 in our analysis). The masses within 200\,kpc can be compared to those from the larger study by \cite{Fox2022} on 74 different clusters, whereas the radius of 400\,kpc is the largest radius shared by all mass maps. \cite{Golubchik2022} also cite masses at two additional radii, corresponding to the Einstein radii derived with their model for source redshifts of $z=1.45$ and $z=2$: M$_{{\rm Golubchik+22,\ 78\,kpc}}=(3.42\pm 0.47)\times$10$^{13}$\,\msun, and M$_{{\rm Golubchik+22,\ 90\,kpc}}=(4.15\pm 0.58)\times$10$^{13}$\,\msun\ respectively.
Our model yields higher masses of 

\begin{figure*}
\begin{center}
	\includegraphics[width=0.75\paperwidth]{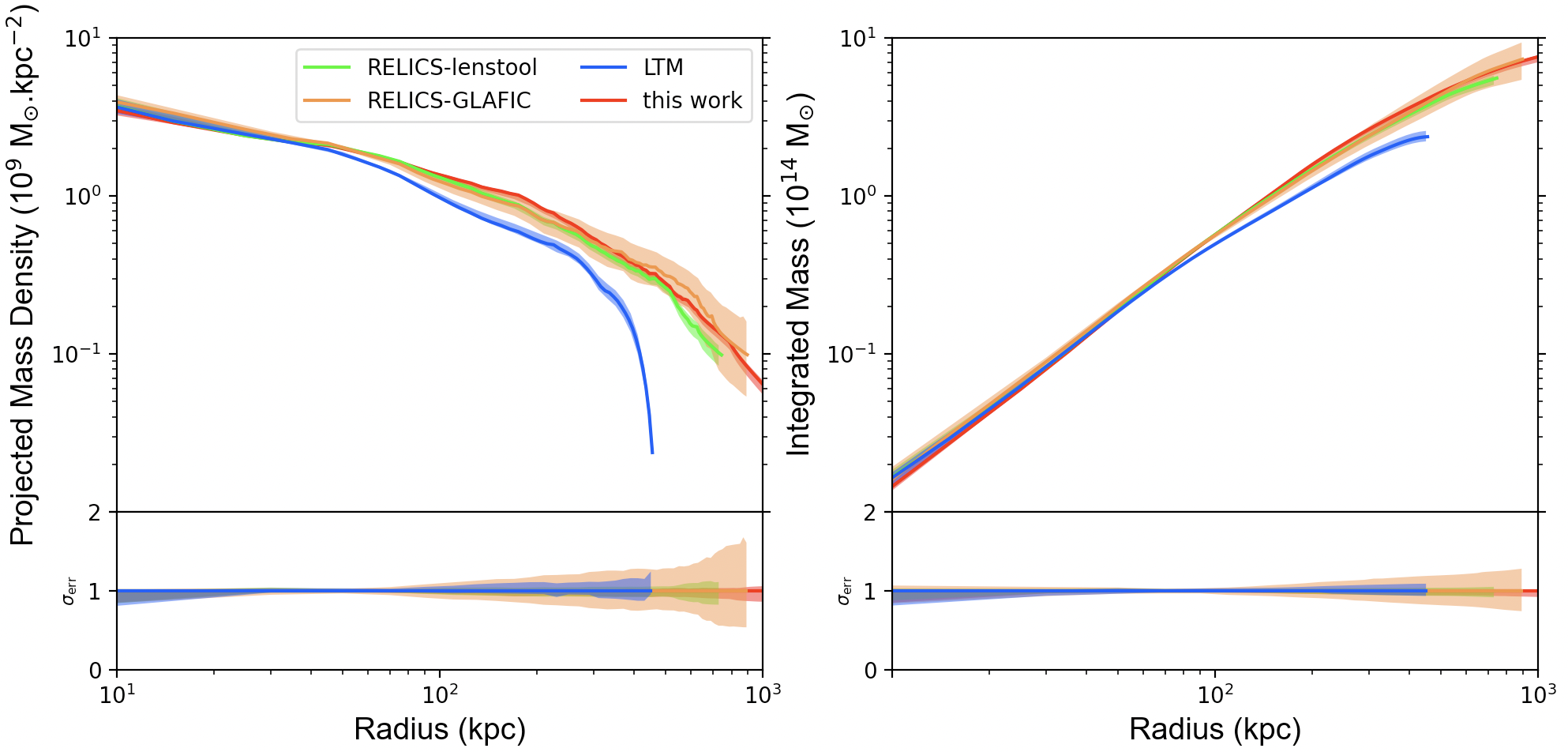}
    \caption{\textit{Left}: Mass-density profiles of SMACS\,J0723 obtained by our analysis (red) and in previous works: RELICS-\textsc{Lenstool} (green), RELICS-GLAFIC (orange), and LTM (blue) with their respective 1$\sigma$ uncertainties (shaded areas). \textit{Right}: Integrated mass profiles obtained for SMACS\,J0723. The graph at the bottom of either panel shows the respective relative 1$\sigma$ uncertainties of each model. As expected, these uncertainties are smallest within the radial range within which most strong-lensing constraints are observed. }
    \label{fig:mass_profil}
\end{center}
\end{figure*}

M$_{{\rm 78\,kpc}}=(3.81\pm 0.02)\times$10$^{13}$\,\msun, and  M$_{{\rm 90\,kpc}}=(4.83\pm 0.03)\times$10$^{13}$\,\msun.
Although these two masses as statistically consistent with each other, the discrepancies may also reflect differences in modeling assumptions and our addition of spectroscopic redshifts. 
The full profile shown in Fig.~\ref{fig:mass_profil} highlights the differences between the various mass profiles. At about 300\,kpc, the mass density for the LTM lens model falls significantly below other measurements.

\begin{table}
	\centering
	\caption{Total enclosed cluster mass at different radii; in units of $10^{12}$ \msun}
	\label{tab:mass}
	\begin{tabular}{l|cccc} % four columns, alignment for each
	Model & M$_{\rm 128\,kpc}$ &  M$_{\rm 200\,kpc}$ &  M$_{\rm 400\,kpc}$ \\ 
	\hline \\[-3mm]
this work & $81.27^{+0.76}_{-0.33}$ & $153.75^{+1.95}_{-0.78}$ & $348.14^{+6.59}_{-3.1}$ \\[2mm]
RELICS-\lenstool\ & $80.8\pm 0.7$ & $146.1\pm 2.1$ & $323^{+8}_{-6}$\\[2mm]
RELICS-GLAFIC & $79.1^{+2.5}_{-1.6}$ & $144.1^{+6.5}_{-5.5}$ & $338^{+26}_{-25}$ \\[2mm]
LTM & $68.6^{+0.5}_{-0.7}$ & $117.3^{+1.1}_{-1.9}$ & $276.9^{+5.8}_{-5.6}$ 
\end{tabular}
\end{table}

We note that \cite{Golubchik2022} report a high RMS uncertainty of 2\farcs3, whereas the RMS of the RELICS model of 0\farcs58 \citep{Sharon2022} is typical for similar cluster lens models based on a fairly limited number of multiple-image systems. By contrast, our new models (which employ many more strong-lensing constraints) yield an RMS of \rms.
This trend is in line with an analysis of simulated clusters \citep{Johnson2016}, which shows that models with a large number of spectroscopic constraints yield more accurate  strong-lensing magnifications and masses.

Following the release of the ERO data, two other teams \citep{Caminha2022,Pascale2022} published lens models for \SMACS. We collaborated with both teams to work toward a set of mutually agreed-upon multiple-image constraints and labels. Here, we present a brief discussion and comparison of the remaining main differences between the three lens models. 

We note that \cite{Caminha2022} present a spectroscopic redshift for system 19 of 1.3825. Due to the low signal-to-noise ratio of the detection and the presence of a skyline on top of the emission, we did not use this redshift as an input constraint in our modeling. We do, however, find a redshift of $1.42^{+0.02}_{-0.02}$ (consistent with theirs) from our fiducial model. As for the RMS of each team's best lens model, \cite{Caminha2022} report $0\farcs51$ and \cite{Pascale2022} quote $0\farcs93$, compared to our value of \rms. Since this work and the analysis by \citet{Caminha2022} use the same modeling software (\lenstool), the difference between our models is due to our inclusion and reliance on a larger (also different) number of constraints and their additional use of an external shear component, while we instead include additional mass components, one of them motivated by the detection of excess ICL.

\citealt{Caminha2022} report an ellipticity for the main cluster-scale halo of 0.51, whereas our comparison model (without ICL clump) has a median ellipticity of 0.86. Although the difference can partly be attributed to differences in the lensing constraints used, we stress that the ellipticity can also be reduced by the external-shear component added in the model of \citealt{Caminha2022}. Our fiducial model (with the ICL clump) presents a lower median ellipticity of 0.67. A more detailed comparison, quantifying, for instance, the influence of each strong-lensing constraint on the model's ellipticity, is beyond the scope of this paper.

\subsubsection{Dynamical state of SMACS J0723}

The distribution of cluster members in SMACS\,J0723 reveals no significant substructure, neither in velocity space (see Section~\ref{sec:clustergal}) nor in projection onto the plane of the sky, suggesting that the mass distribution is adequately described by a single cluster-scale component. However, in order to recover the geometry of multiple images newly discovered in the \JWST\ observations (i.e., to minimize the RMS of our model), we require a more sophisticated mass model that incorporates two additional diffuse mass concentrations as discussed in Section~\ref{sec:diffmass}. These could be interpreted as remnant/tracers of past dynamical activity in the cluster. We stress that our final mass model, which includes the aforementioned additional components, has an RMS of \rms, a substantial improvement over the value of 1\farcs26 for a model which only includes a single cluster-scale halo centered on the BCG. 

As discussed in Section \ref{sec:clustergal}, the distribution of the radial velocities of the cluster galaxies does not show compelling evidence of substructure along the line of sight. However, the offset between the radial velocity of the BCG and the centroid of the overall redshift distribution suggests that SMACS\,J0723 is not fully relaxed, an assessment that is supported by the complex mass distribution required and obtained from our strong-lensing analysis.

We report a large ellipticity of 0.86 for our comparison model (without ICL clump). By contrast, our fiducial model (with ICL clump) features an ellipticity of only 0.67. The fact that the addition of a mass component associated with the ICL reduces the overall ellipticity lends further support to the interpretation that the cluster is not a relaxed. Some previous studies have also used external shear to motivate an additional mass component \citep{Mahler2018} which also affects the ellipticity. Since the impact and interplay between components in the context of cluster-relaxation assessments remains an active area of exploration \citep{Zitrin2015,Desprez2018,Lagattuta2019,Ghosh2021}, we defer a more in-depth investigation of the cluster state to future work.

Additional evidence for dynamic activity and ongoing cluster evolution is provided by the presence of the excess ICL itself shown in Fig.~\ref{fig:icl}. As discussed in Section~\ref{sec:diffmass}, these ICL features play an important role for our lens modeling efforts: without the presence of ICL revealed by the \textit{JWST} ERO data, refinements to our mass model in the west and south-west regions would have been driven solely by statistics, i.e., the need to lower the RMS, rather than being supported and motivated by physical evidence for the presence of mass in these regions of SMACS\,J0723.

\begin{table}
	\centering
	\caption{Surface area $\sigma_{\mu}$ in the source plane with magnifications in excess of a given magnification $\mu$ for this work and the RELICS-lenstool published in \citealt{Sharon2022}}. We quote $\sigma_{\mu}(>\mu)$ for $\mu=3$, 5, and 10 for a source at redshift $z=9$. 
	\label{tab:magnification}
	\begin{tabular}{l|cccc} % four columns, alignment for each
	Model & $\sigma_{\mu}(3)$ &  $\sigma_{\mu}(5)$ &  $\sigma_{\mu}(10)$ \\ 
	\hline \\[-3mm]
this work & 1.52 &  1.0	 & 0.7	  \\
RELICS-\lenstool\ & 1.5 & 0.95 & 0.5
\end{tabular}
\end{table}

\subsection{Magnification measurements}

\begin{figure}
\begin{center}
	\includegraphics[width=\columnwidth]{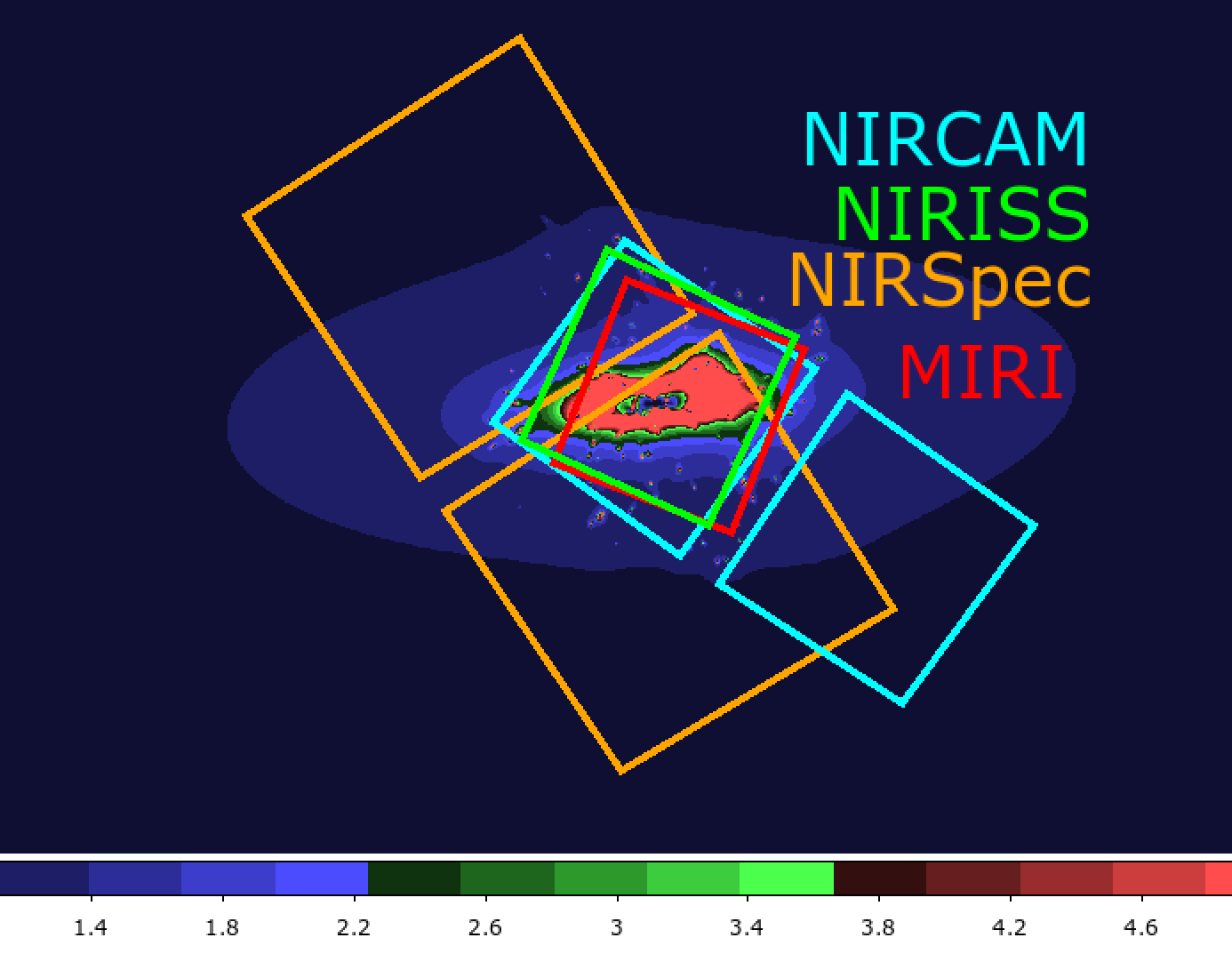}
    \caption{Magnification map obtained from our mass model for a source at redshift $z=9$. Overlaid are the footprints of \JWST's instruments.}
    \label{fig:magni}
    \end{center}
\end{figure}

Thanks to the dramatically increased number of multiple-image systems uncovered with JWST, as well as the availability of partial spectroscopic coverage to anchor the mass and shape of the cluster lens, we are able to derive magnification maps for SMACS\,J0723 across the footprints of all \JWST\ instruments. Fig.~\ref{fig:magni} shows the magnification map obtained for sources at redshift $z=9$.

Following the method presented by \cite{Wong2012} and subsequently applied to HFF analyses \citep[e.g.][]{Jauzac2014,Jauzac2015b,Lam2014,Wang2015,Hoag2016}, we use the surface area in the source plane, $\sigma_{\mu}$, above a given magnification factor $\mu$ as a metric to quantify the efficiency of the lens to magnify high-redshift background galaxies, noting that $\sigma_{\mu}$ is directly proportional to the unlensed comoving volume covered at high redshift at a given magnification $\mu$. 

\begin{figure}
\begin{center}
	\includegraphics[width=\columnwidth]{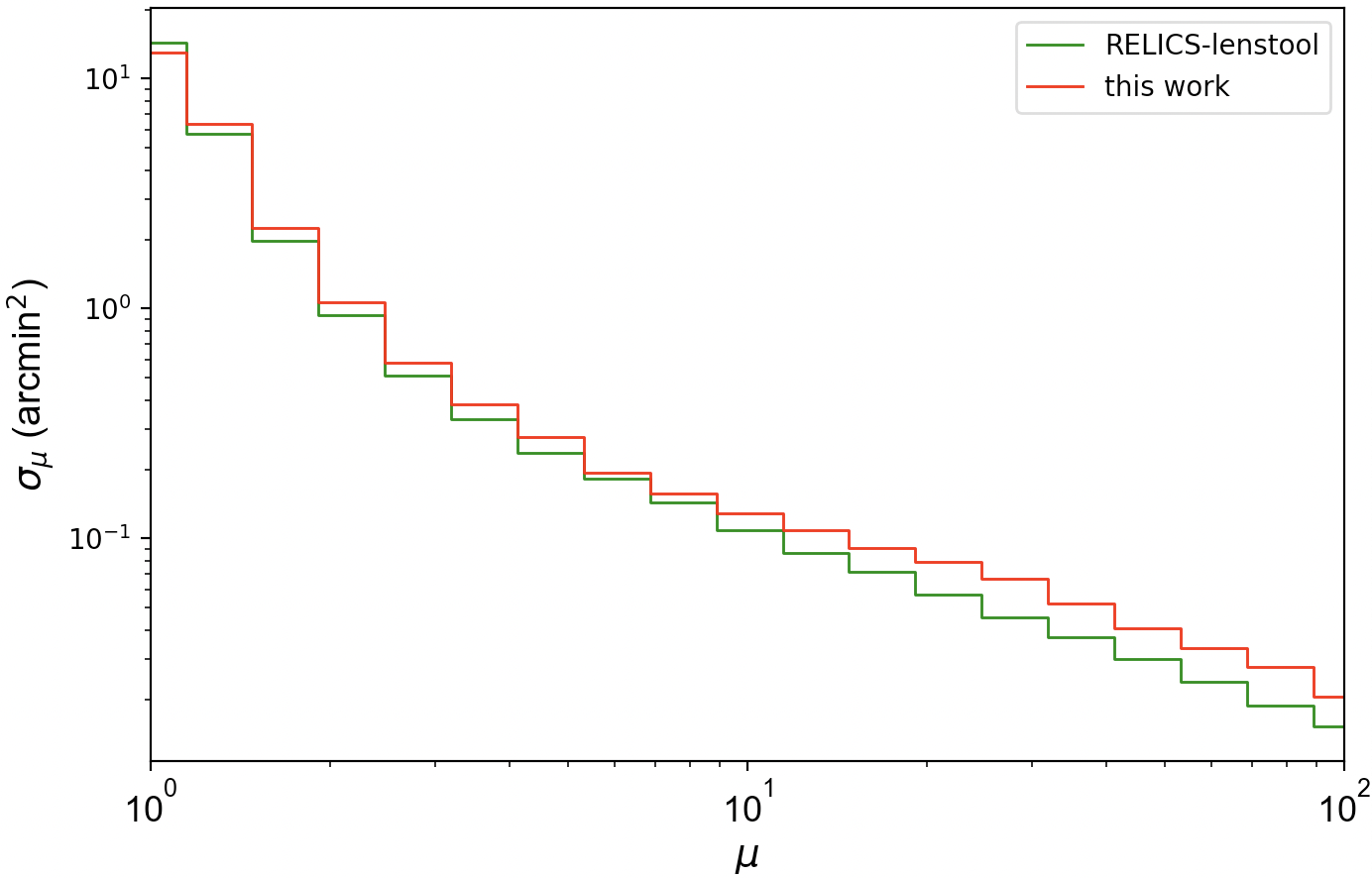}
    \caption{Surface area in the source plane within a 300\arcsec\ box centered on the cluster at a magnification above a given threshold $\mu$ for a source at z$=$9. We here compare the values obtained with our updated mass model with those from the RELICS-\textsc{Lenstool} model.}
    \label{fig:magni_histo}
\end{center}
\end{figure}

Fig.\,\ref{fig:magni_histo} shows the evolution of $\sigma_{\mu}({>}\mu)$ as a function of the magnification obtained from our final mass model of SMACS\,J0723 for a source at a redshift $z=9$. Our model yields $\sigma_{\mu} (\mu>3) = 1.52$\,arcmin$^{2}$, $\sigma_{\mu} (\mu>5) = 1.0$\,arcmin$^{2}$, and $\sigma_{\mu} (\mu>10) = 0.7$\,arcmin$^{2}$. Table\,\ref{tab:magnification} compares these values with measurements obtained by RELICS-\lenstool, i.e., with the same mass-modeling algorithm. 
The pre-\JWST\ RELICS-\lenstool\ model used seven unique systems as constraints, with no spectroscopic redshifts (private communication), and yielded smaller areas than found from the model presented in this paper, especially at very high magnifications, suggesting that SMACS\,J0723 is a more powerful cluster lens than initially believed.  

As a general caveat regarding magnification maps, we acknowledge limitations caused by a lack of constraints at large cluster-centric distances. We note, however, that the wide-angle X-ray observation performed with \textit{Chandra}'s ACIS-I detector (discussed in Sections~\ref{sec:xrays} and \ref{sec:xres}) does not reveal any further sources indicative of gravitationally collapsed mass concentrations in the vicinity of SMACS\,J0723. 
We therefore consider our magnification maps (and the associated error maps) to be robust and make them available to the community as part of this publication. \review{We acknowledge that magnification values beyond the region where multiple images reside result from a model extrapolation and could be affected by systematic uncertainties}

\subsection{Intra-cluster medium (ICM)}
\label{sec:xres}
\subsubsection{X-ray morphology}

Fig.~\ref{fig:xopt} shows iso-intensity contours of the adaptively smoothed X-ray surface brightness from SMACS\,J0723 overlaid on the \JWST\ color image of the cluster core. We find the X-ray emission to feature a well defined, single peak at a location that coincides perfectly with that of the BCG\footnote{Although a direct astrometric alignment of the \JWST\ and \textit{Chandra} images is precluded by the fact that all X-ray point sources detected in the \textit{Chandra} observations fall outside the \JWST\ field of view, a comparison with wide-field J-band imaging obtained by the VISTA Hemisphere Survey (ESO Progamme 179.A-2010, PI McMahon) limits the relative astrometric misalignment to about 1\arcsec.}. While such alignment can be viewed as a sign of a system in dynamic equilibrium, the clearly disturbed X-ray morphology outside the very core region represents unambiguous evidence of recent merger activity.

\begin{figure*}
\centering
	\includegraphics[width=\textwidth]{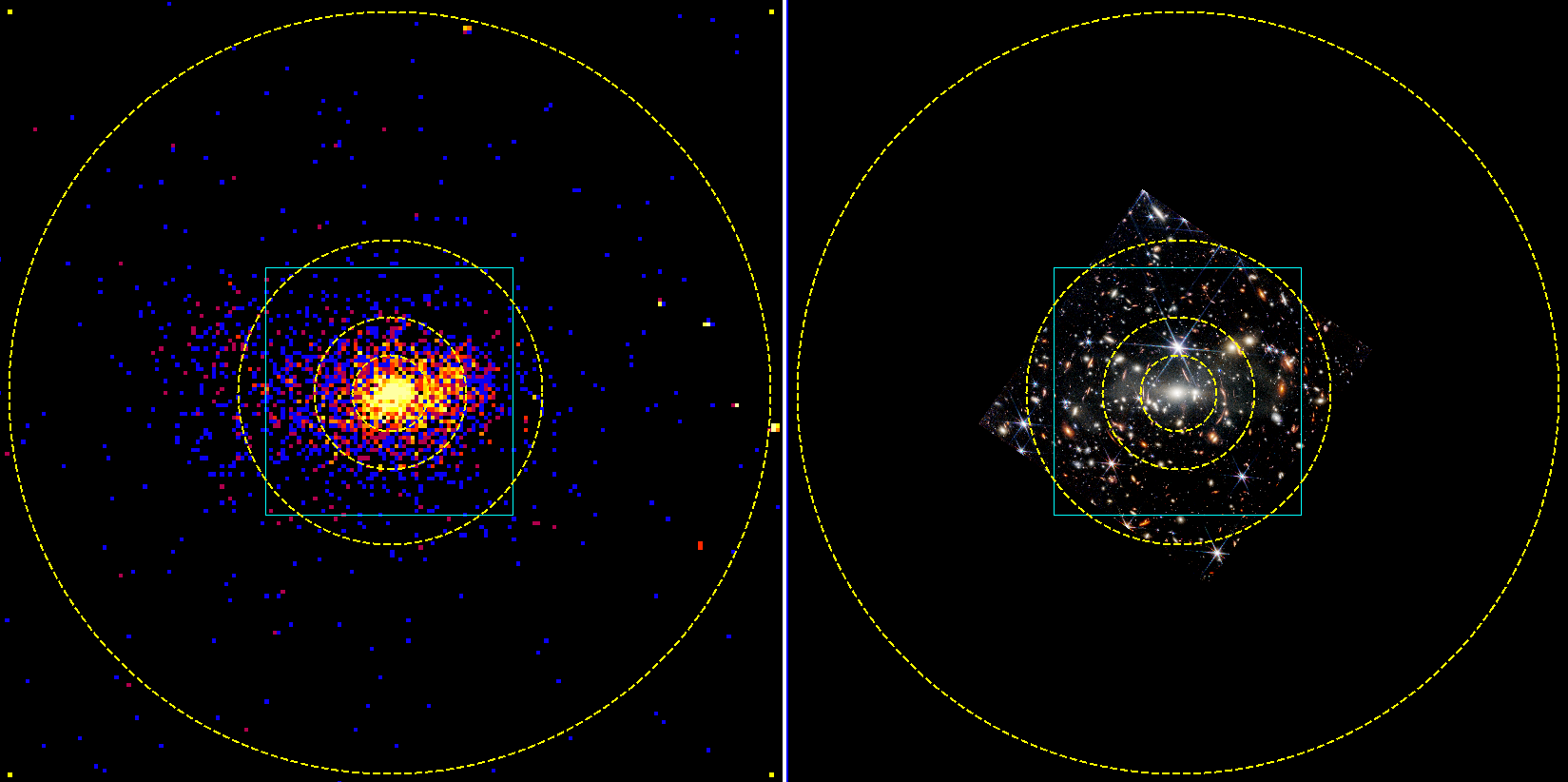}
    \caption{Regions of interest for our measurements of the ICM temperature overlaid on the \textit{Chandra} ACSI-I image of SMACS\,J0723 (left; 2\arcsec pixels, 0.5--7 keV, logarithmic intensity scaling) and on the \JWST\ image of the system (right). The dashed circles have radii of 100, 200, 400, and 1000\,kpc, respectively, at the cluster redshift. The cyan square marks the region shown in Fig.~\ref{fig:xopt}.}
    \label{fig:xoptreg}
\end{figure*}

\subsubsection{ICM temperature}

\begin{figure}
\centering
    \includegraphics[width=\columnwidth]{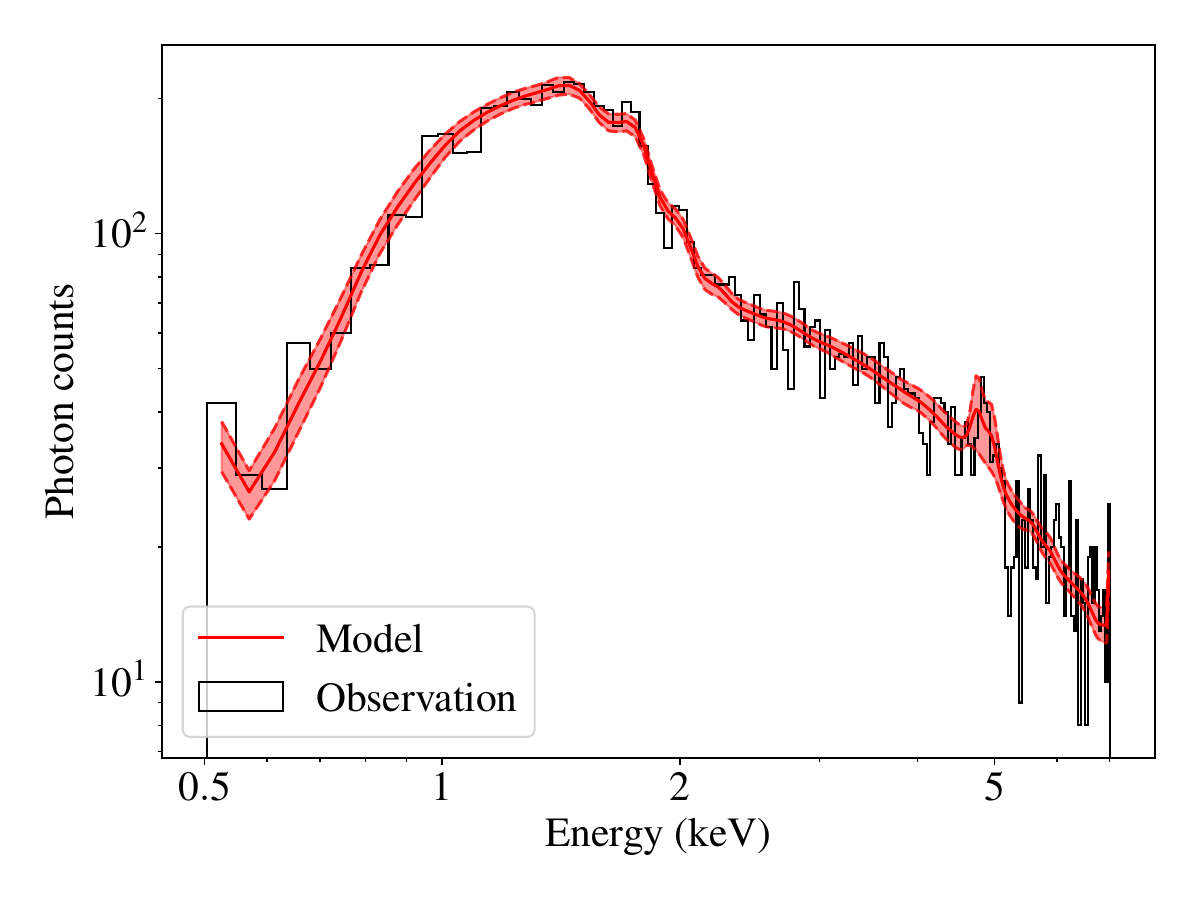}
    \caption{Global spectrum of the observed ICM emission within 1\,Mpc of the X-ray peak, \review{corresponding to approximately r$_{\rm 1000}$, (the radius enclosing a thousand times the mean density of the universe at that redshift)}. Overlaid in red is the best-fit APEC model with its associated $68$\% confidence range.}
    \label{fig:xspec}
\end{figure}

The spectral analysis summarized in Section~\ref{sec:xrays} yields a global ICM temperature (within 1 Mpc of the X-ray peak) of k$T=9.80^{+1.54}_{-1.37}$\,keV, a metallicity of $Z=0.38^{+0.12}_{-0.11}$\,$Z_{\odot}$, and an equivalent hydrogen column density of $n_H=1.94^{+0.03}_{-0.03}\times10^{21}$ cm$^{-2}$. Fig.~\ref{fig:xspec} shows the global spectrum as well as the best-fit spectral model with $68$ per cent uncertainties (as represented by the associated subset of the sampled parameter distributions). Our best-fit value for $n_H$ agrees to better than $1\sigma$ with the total hydrogen (i.e., HI and HII) column density of $2.21\times10^{21}$ cm$^{-2}$ measured by \citet{2013MNRAS.431..394W}.

\begin{figure}
\centering
    \includegraphics[width=\columnwidth]{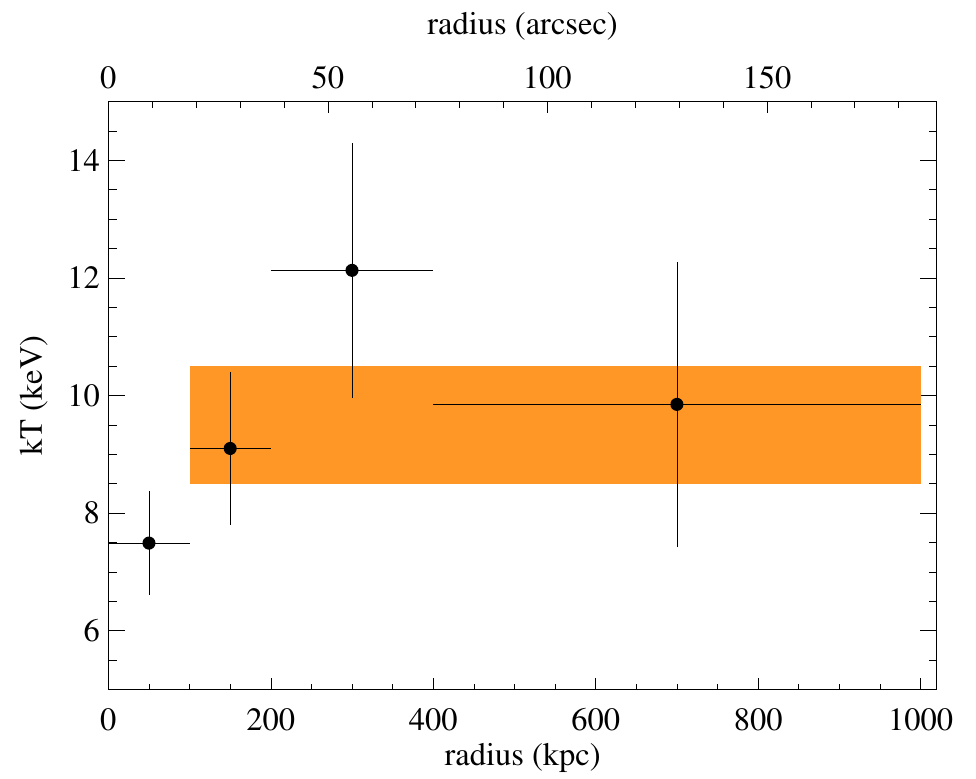}
    \caption{ICM temperature measurements within the regions shown in Fig.~\ref{fig:xoptreg}; vertical bars represent $1\sigma$ uncertainties, horizontal bars represent the width of the respective annulus. The ambient ICM temperature in the combined regions beyond $r=100$ kpc (i.e., within the annulus from 200 to 1000\,kpc) with its $1\sigma$ error is shown as an orange rectangle.}
    \label{fig:ktprof}
\end{figure}

We attempt to constrain spatial variations in the ICM temperature by fitting separate spectral models to the data in the regions marked in Fig.~\ref{fig:xoptreg}. Acknowledging the reduced signal in these smaller regions, we adopt the Galactic total $n_{\rm H}$ value; we also freeze the metallicity at $Z=0.3$ for these fits, in agreement with typical metal-abundance values observed for non-relaxed clusters at similar redshift \citep{2015A&A...578A..46E}. The results, shown in Fig.~\ref{fig:ktprof}, are consistent with a constant ICM temperature but suggest (at less than 2$\sigma$ significance) a slight drop in k$T$ in the very core of SMACS\,J0723.

\subsubsection{Gas mass}

\begin{figure*}
\centering
    \begin{minipage}{0.49\linewidth}
        \includegraphics[width=\columnwidth]{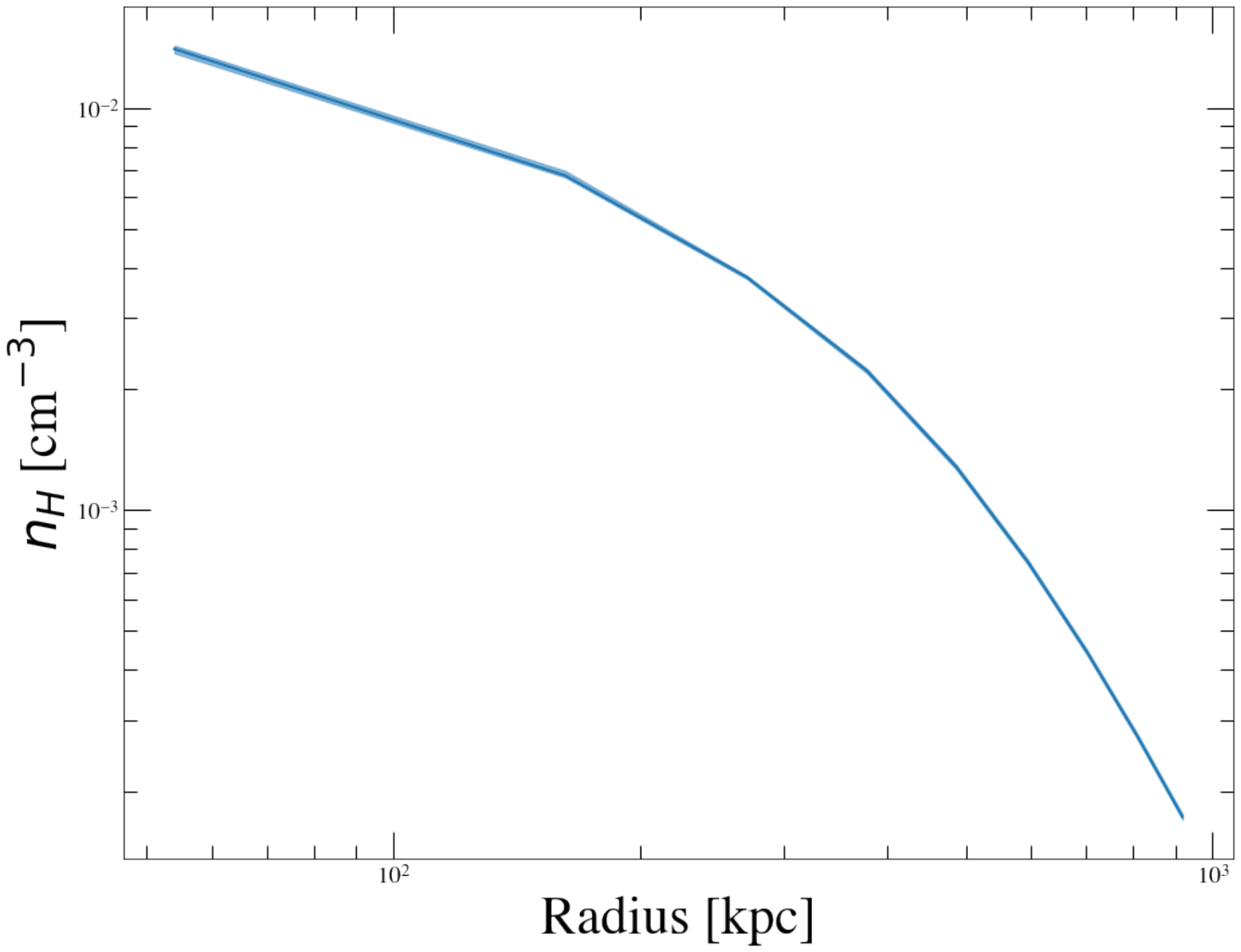}
    \end{minipage} \hfill
	\begin{minipage}{0.49\linewidth}
    \includegraphics[width=\columnwidth]{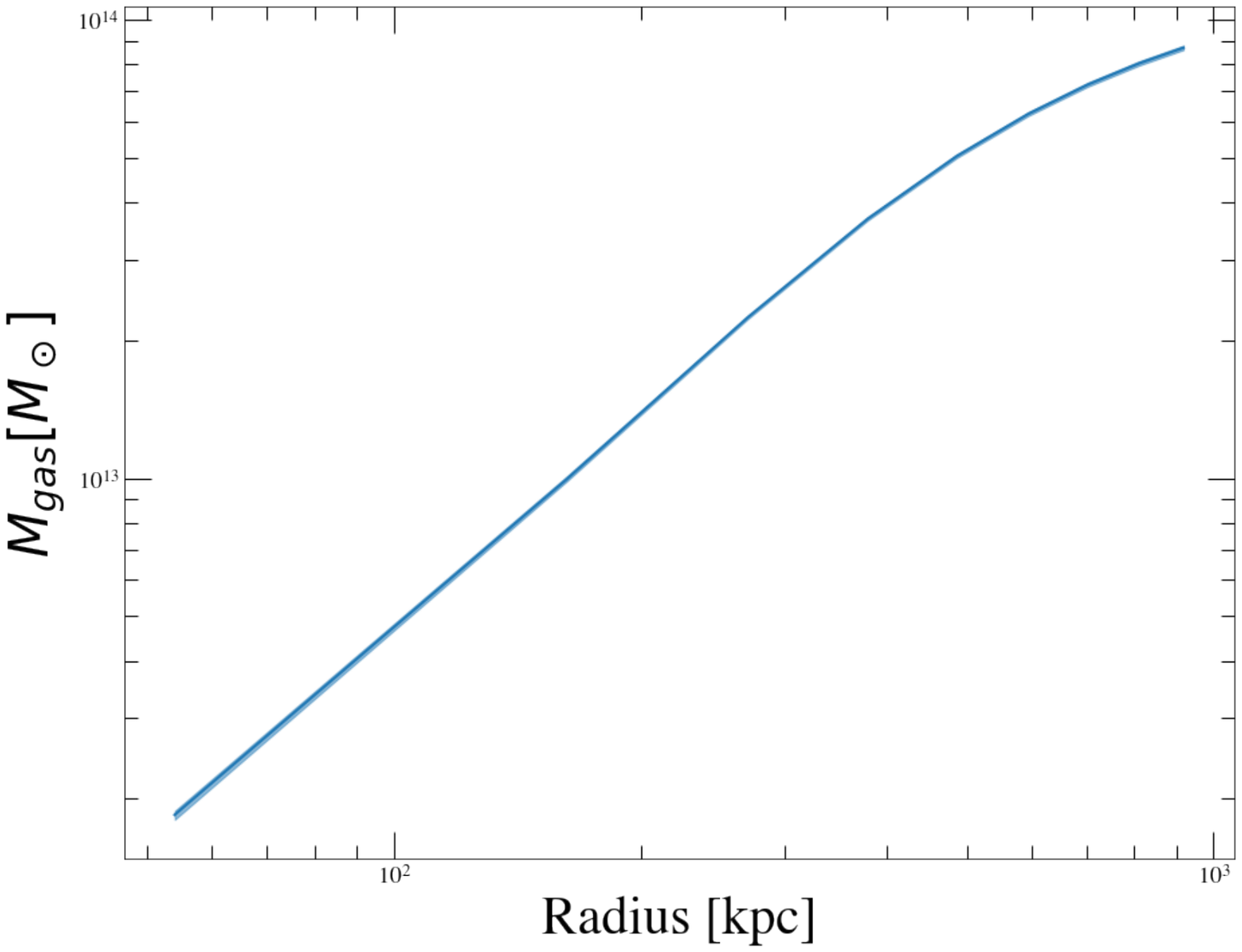}
    \end{minipage}
    \caption{Profiles of the ICM density (left) and the cumulative gas mass (right) as determined from a spherical-deprojection analysis.}
    \label{fig:gasprof}
\end{figure*}

We perform a multi-scale deprojection of the gas density and gas mass using the \textsc{pyproffit} python package developed by \citet{2020OJAp....3E..12E}. The analysis uses counts and background maps in the $0.5$-$2.0$ keV energy band, an associated monochromatic exposure map for an energy of $1.2$ keV, as well as the values from our best-fit spectral model. The resulting profiles of the ICM density and the cumulative gas mass are shown in Fig.~\ref{fig:gasprof} and place the total gas mass of SMACS\,J0723 at almost $10^{14}$ \msun. A comparison with the total gravitational mass derived from our lens model (Fig.~\ref{fig:mass_profil} and Table~\ref{tab:mass}) yields a gas-mass fraction of just under 10\% for the cluster core, typical of massive clusters in general. A more detailed investigation of, e.g., the baryon fraction across the system would require a significantly deeper X-ray observation and much more sophisticated spatial modeling of the ICM.

\subsubsection{Global properties}

For reference, we summarize all global cluster properties derived for SMACS\,J0723 from the only existing, dedicated X-ray observation of the cluster in Table~\ref{tab:xprop}. All values are computed from the emission within $r=1$ Mpc which is very close to $r_{\rm 1000}$.

With a total X-ray luminosity well in excess of $10^{45}$ erg s$^{-1}$ in the ROSAT energy band (0.1--2.4 keV) in which the system was originally discovered (see Section~\ref{sec:target}), the properties of SMACS\,J0723 established here are a testament to the power of X-ray selection of clusters in general, and of the MACS project in particular, to uncover exceptionally massive clusters that stand to advance our understanding of a broad range of science topics, from cluster formation and evolution to lensing-assisted, ever-deeper views of the distant Universe.

\begin{table}
    \centering
    \begin{tabular}{cc}
        R.A.\ Decl. (J2000)& k$T$ (keV) \\
        \hline \\[-3mm]
        %pyproffit results
        %$7:23:18.04$ $-73:27:19.79$& $9.80^{+1.54}_{-1.37}$& $18.15^{0.27}_{-0.25}$ &$17.48^{0.26}_{-0.24}$& $29.02^{0.43}_{-0.40}$&$55.21^{0.82}_{-0.77}$&$37.14^{0.55}_{-0.52}$\\
        %Sherpa results
        07:23:18.0 \hspace*{3mm}$-$73:27:19 & $9.8^{+1.5}_{-1.4}$ \\ \\
%    \hline\hline
    \end{tabular}
    \begin{tabular}{ccc}
        energy band & $f_{\rm X}$ ($10^{-13}$ erg s$^{-1}$ cm$^{-2}$)  & $L_{\rm X}$ ($10^{44}$ erg s$^{-1}$)  \\
        \hline \\[-3mm]
          0.1--2.4 keV & $32.1^{+1.2}_{-1.1}$ & $18.6^{+0.7}_{-0.6}$\\[2mm]
          0.5--2.0 keV & $23.2^{+0.9}_{-0.8}$ & $13.4\pm 0.5$\\[2mm]
          0.5--7.0 keV & $58.2\pm 1.3$ & $33.6\pm 0.8$\\[2mm]
          2--10 keV    & $43.6^{+2.5}_{-2.6}$ & $25.2^{+1.4}_{-1.5}$\\[2mm]
          bolometric   & $73.3\pm 2.3$ & $42.4^{+1.4}_{-1.3}$\\[2mm]

       %pyproffit results
       %$11.15^{0.16}_{-0.16}$ &$10.73^{0.16}_{-0.15}$ & $17.82^{0.26}_{-0.25} $&$33.90^{0.50}_{-0.47}$ &$2.80^{0.34}_{-0.32}$
       %Sherpa results

    \end{tabular}
    \caption{Global X-ray properties of SMACS\,J0723 computed within $r=r_{\rm 1000}$. Unabsorbed fluxes and total luminosities are both point-source corrected.}
    \label{tab:xprop}
\end{table}

\section{Discussion and Conclusion}
\label{sec:conc}

We create and make available to the scientific community a robust strong-lensing mass model of the galaxy cluster SMACS\,J0723 at $z=0.39$, the first strong-lensing cluster to be observed with \jwst. Our model uses \JWST\ ERO data, as well as archival, multi-wavelength data of the cluster, from optical to X-ray wavelengths, and combines both imaging and spectroscopic observations. We identify 17 new multiple-image systems.
%, increasing the total number of known strongly lensed galaxies in this field to 21. 
We report a total number of 30 candidate multiple-image systems, two of which are isolated galaxy-galaxy lensed sources. Of the final 28 cluster-wide multiple-image systems, we use 21, namely 19 robust systems and two additional candidates located near the intra-cluster light concentrations identified by us.
\review{Our best-fit mass model contains only one large cluster-scale halo and includes one diffuse large-scale halo that accounts for mass traced by the cluster ICL.  Additional halos have masses closer to galaxy-scale halos. These halos bring flexibility to our model to adjust their nearby multiple-image systems. As a result, our model is able to reproduce overall} the positions of the strong-lensing features to within \rms\ (RMS).

The mentioned excess stellar cluster light (low surface-brightness features that appear clearly on large scales and are enhanced in the \textit{JWST} imaging by median filtering) may represent the signature of a recent merger event. Indeed, the combined evidence from our analysis of the overall mass distribution, radial velocities of cluster galaxies, and ICM properties also suggests that SMACS\,J0723 recently underwent a merger along an axis close to our line of sight but is well on its way to relaxation, as reflected in the nearly perfect alignment of the X-ray peak with the BCG and the overall mass distribution, as well as the increased ICM cooling in an emerging compact gaseous cluster core.

By combining greatly increased sensitivity with broad spectral coverage and spectacular spectroscopic capabilities, \jwst's observation of SMACS\,J0723 reveals exquisite panchromatic details that not only dramatically facilitate the identification of multiple images of galaxies at redshift greater than 5 but also provide additional leverage to constrain the dynamical and merger history of clusters. 

\section*{Acknowledgements}
% hst data acknowledgement
Based on observations made with the NASA/ESA \textit{Hubble Space Telescope}, obtained at the Space Telescope Science  Institute, which is operated by the Association of Universities for Research in Astronomy, Inc., under NASA contract NAS 5-26555. These observations are associated with programs GO-11103, GO-12166, GO-12884, and GO-14096.

%jwst data acknowledgement
This work is based on observations made with the NASA/ESA/CSA \textit{James Webb Space Telescope}. The data were obtained from the Mikulski Archive for Space Telescopes at the Space Telescope Science Institute, which is operated by the Association of Universities for Research in Astronomy, Inc., under NASA contract NAS 5-03127 for \textit{JWST}. These observations are associated with program \#2736. We thank Ian Smail for insightful discussions.
GM acknowledges funding from the European Union’s Horizon 2020 research and innovation programme under the Marie Skłodowska-Curie grant agreement No MARACHAS - DLV-896778. MJ is supported by the United Kingdom Research and Innovation (UKRI) Future Leaders Fellowship `Using Cosmic Beasts to uncover the Nature of Dark Matter' (grant number MR/S017216/1). HE gratefully acknowledges support from STScI grants GO-12166 and GO-12884.  Durham authors acknowledge STFC support through ST/T000244/1.
HA acknowledges support from CNES. PN acknowledges the Black Hole Initiative (BHI) at Harvard University, which is supported by grants from the Gordon and Betty Moore Foundation and the John Templeton Foundation, for hosting her.

%%%%%%%%%%%%%%%%%%%%%%%%%%%%%%%%%%%%%%%%%%%%%%%%%%
%\section*{Data Availability}

 % Data Availability Statements provide a standardised format for readers to understand the availability of data underlying the research results described in the article. The statement may refer to original data generated in the course of the study or to third-party data analysed in the article. The statement should describe and provide means of access, where possible, by linking to the data or providing the required accession numbers for the relevant databases or DOIs.

%%%%%%%%%%%%%%%%%%%% REFERENCES %%%%%%%%%%%%%%%%%%

\bibliographystyle{yahapj}
\bibliography{example}

%%%%%%%%%%%%%%%%%%%%%%%%%%%%%%%%%%%%%%%%%%%%%%%%%%

%%%%%%%%%%%%%%%%% APPENDICES %%%%%%%%%%%%%%%%%%%%%

\appendix

\section{Multiple images}
%\begin{comment}
\begin{figure}
\centering
    \includegraphics[width=0.15\columnwidth]{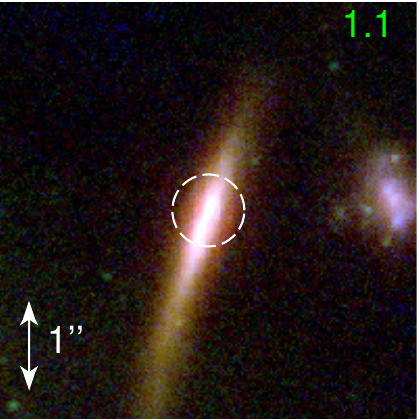}
    \includegraphics[width=0.15\columnwidth]{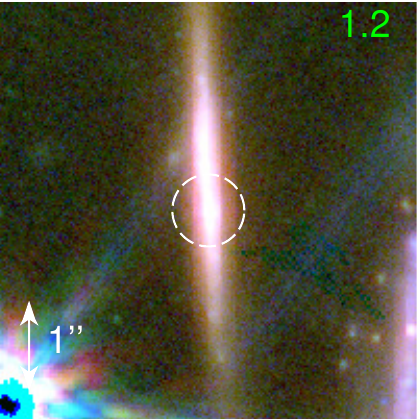}
    \includegraphics[width=0.15\columnwidth]{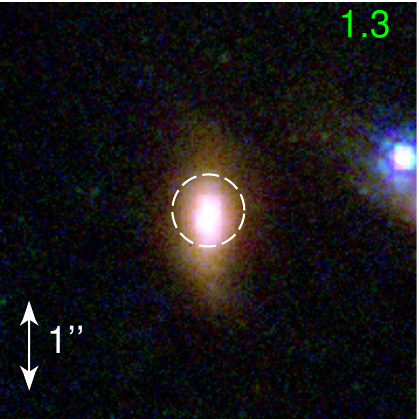}
    \includegraphics[width=0.15\columnwidth]{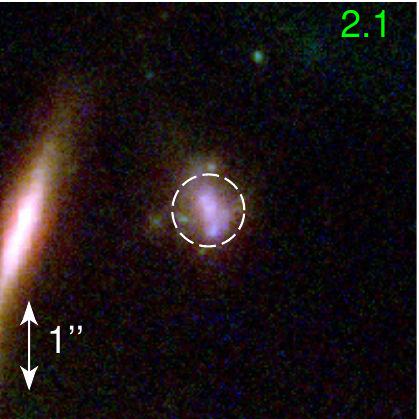}
    \includegraphics[width=0.15\columnwidth]{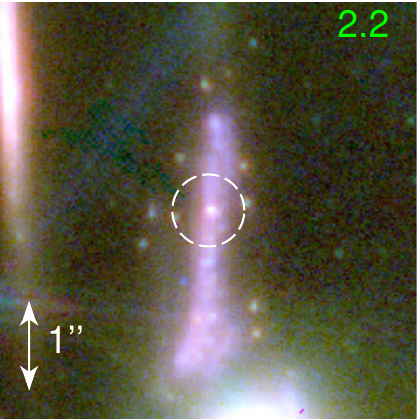}
    \includegraphics[width=0.15\columnwidth]{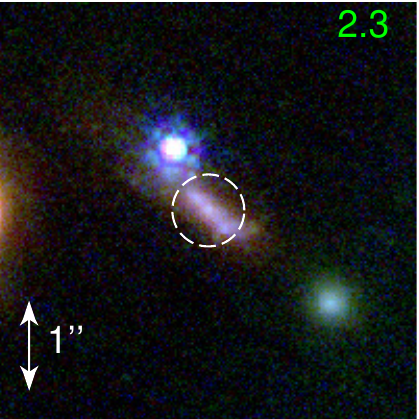}
    \includegraphics[width=0.15\columnwidth]{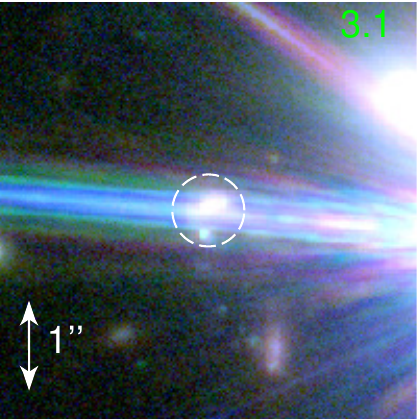}
    \includegraphics[width=0.15\columnwidth]{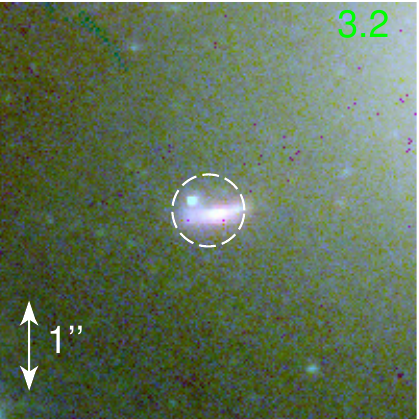}
    \includegraphics[width=0.15\columnwidth]{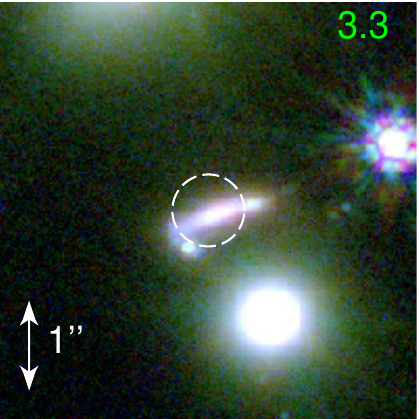}
    \includegraphics[width=0.15\columnwidth]{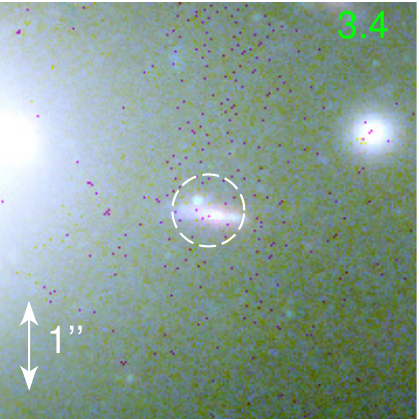}
    \includegraphics[width=0.15\columnwidth]{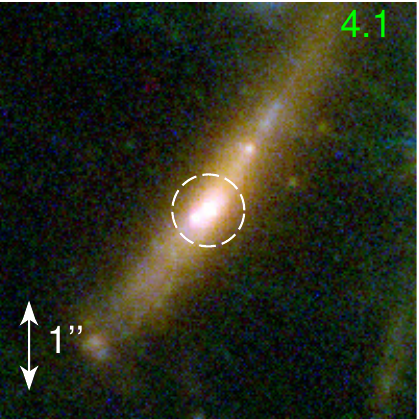}
    \includegraphics[width=0.15\columnwidth]{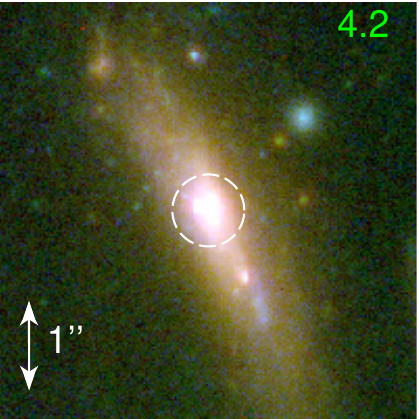}
    \includegraphics[width=0.15\columnwidth]{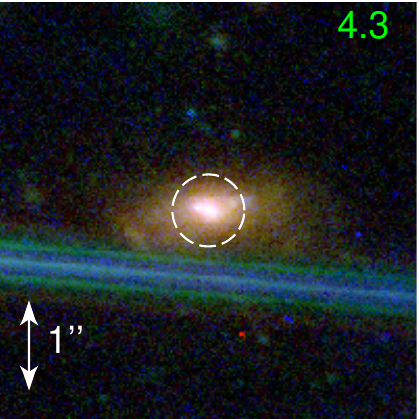}
    \includegraphics[width=0.15\columnwidth]{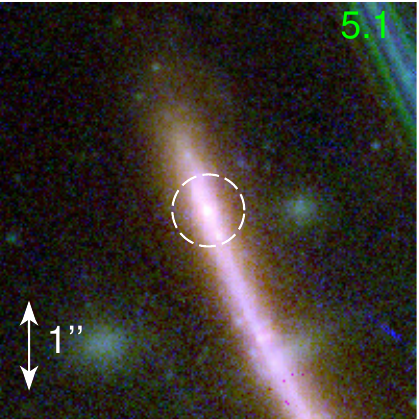}
    \includegraphics[width=0.15\columnwidth]{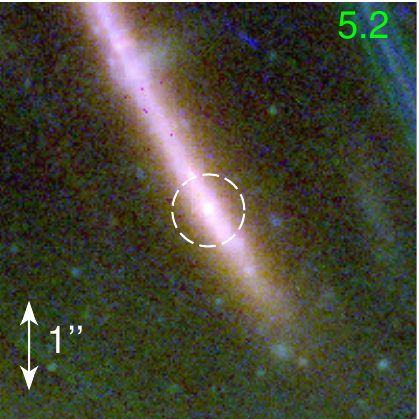}
    \includegraphics[width=0.15\columnwidth]{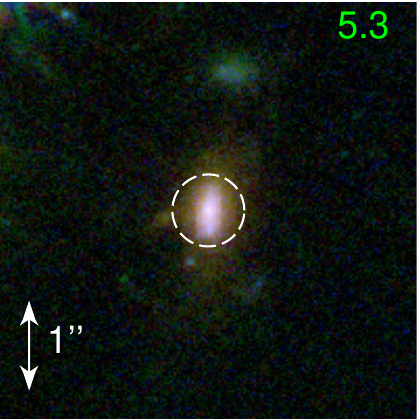}
    \includegraphics[width=0.15\columnwidth]{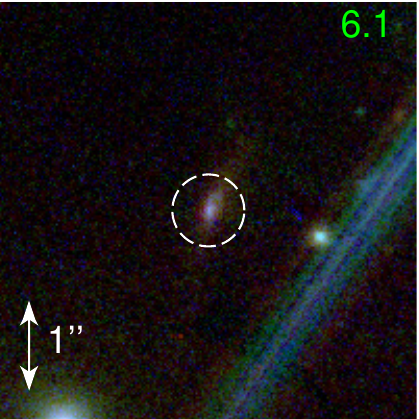}
    \includegraphics[width=0.15\columnwidth]{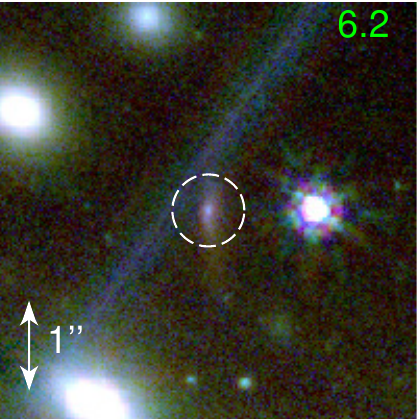}
    \includegraphics[width=0.15\columnwidth]{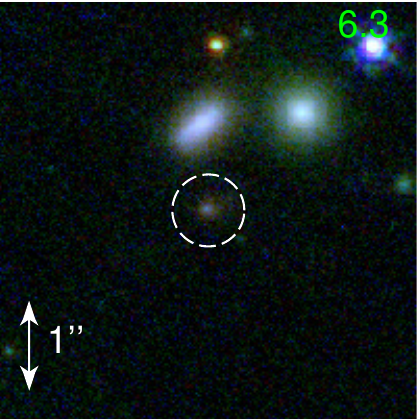}
    \includegraphics[width=0.15\columnwidth]{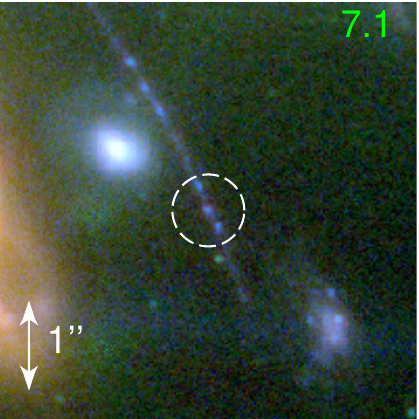}
    \includegraphics[width=0.15\columnwidth]{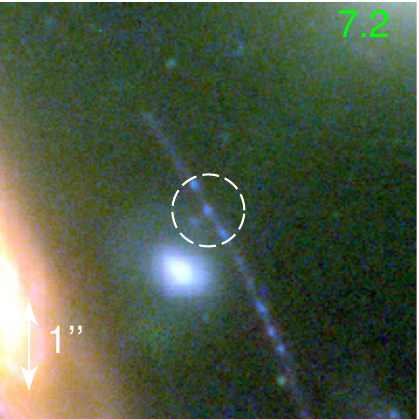}
    \includegraphics[width=0.15\columnwidth]{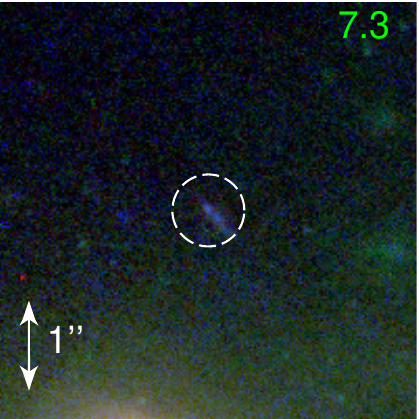}
    \includegraphics[width=0.15\columnwidth]{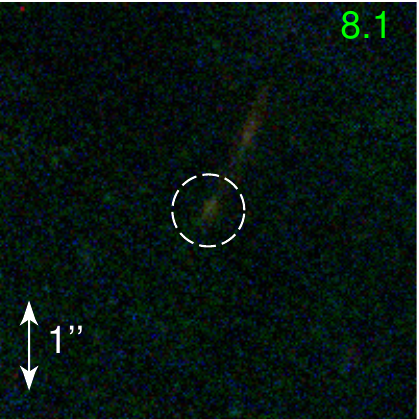}
    \includegraphics[width=0.15\columnwidth]{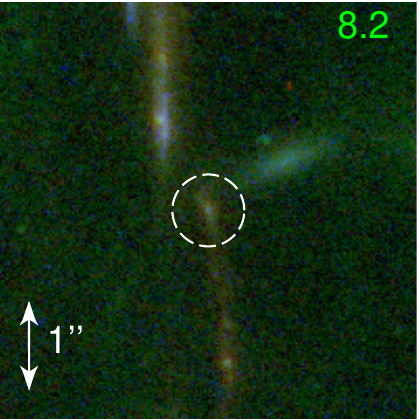}
    \includegraphics[width=0.15\columnwidth]{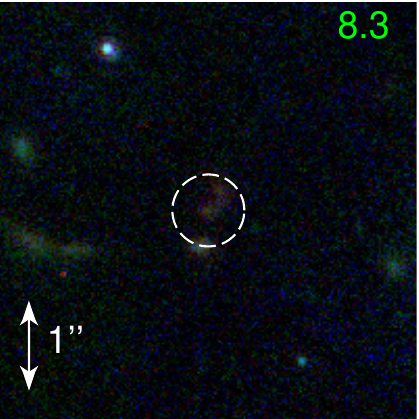}
    \includegraphics[width=0.15\columnwidth]{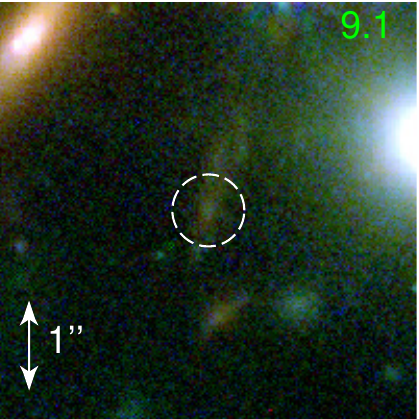}
    \includegraphics[width=0.15\columnwidth]{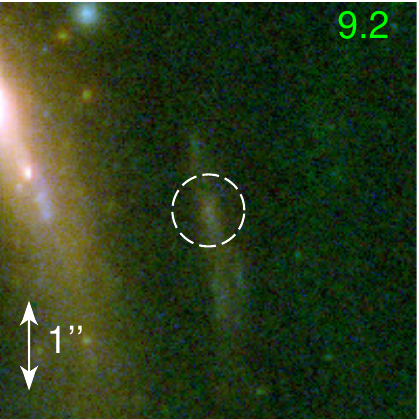}
    \includegraphics[width=0.15\columnwidth]{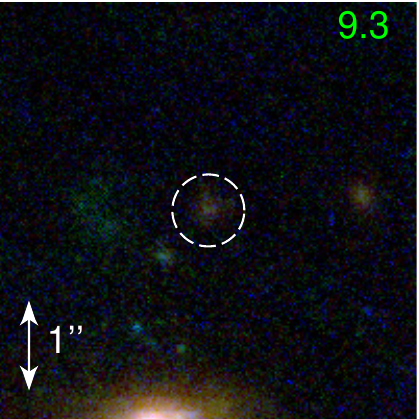}
    \includegraphics[width=0.15\columnwidth]{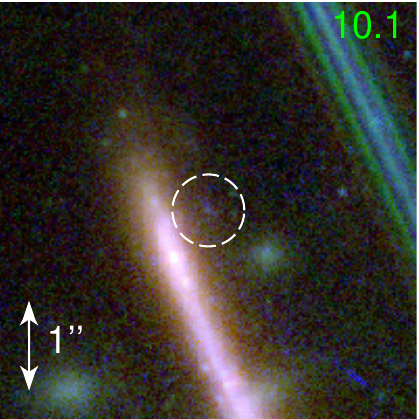}
    \includegraphics[width=0.15\columnwidth]{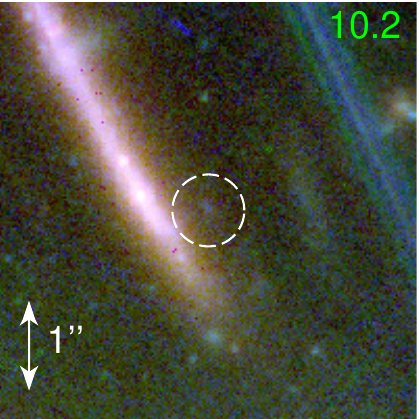}
    \includegraphics[width=0.15\columnwidth]{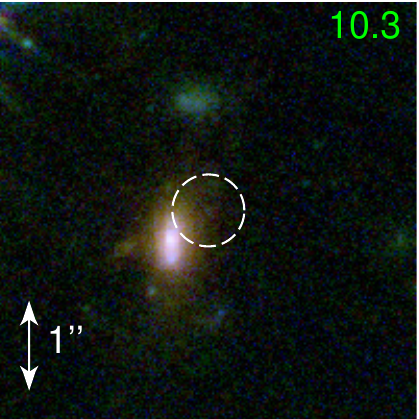}
    \includegraphics[width=0.15\columnwidth]{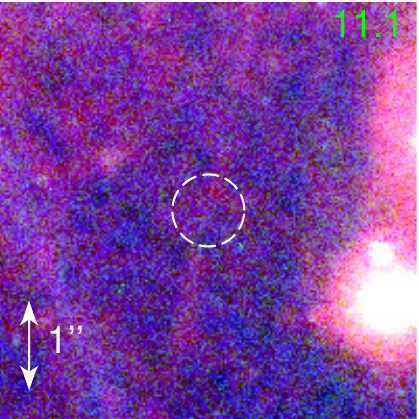}
    \includegraphics[width=0.15\columnwidth]{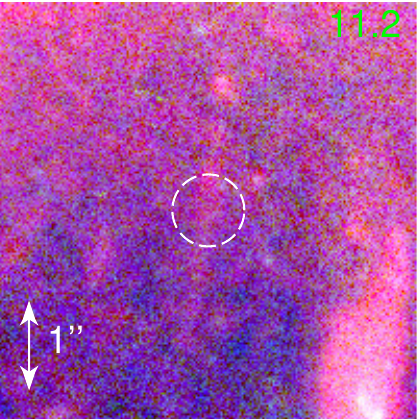}
    \includegraphics[width=0.15\columnwidth]{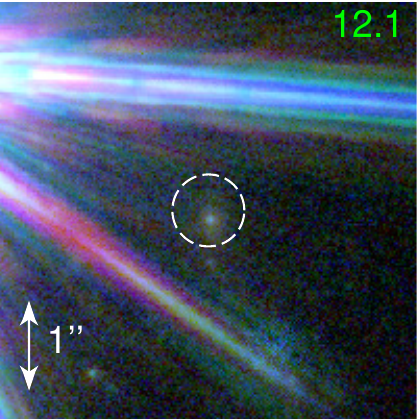}
    \includegraphics[width=0.15\columnwidth]{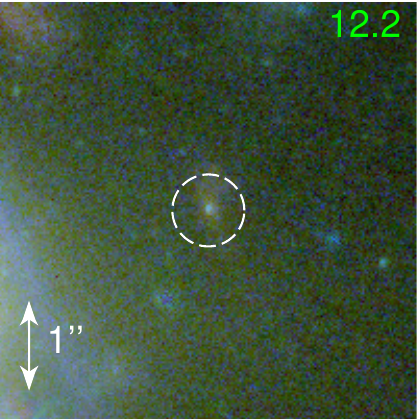}
    \includegraphics[width=0.15\columnwidth]{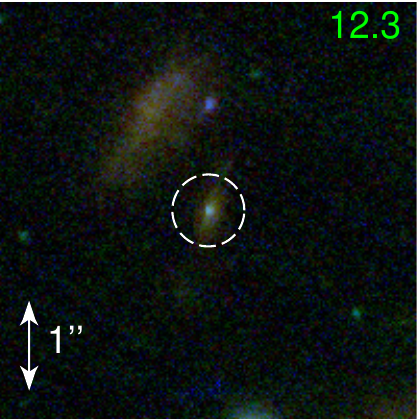}
    \includegraphics[width=0.15\columnwidth]{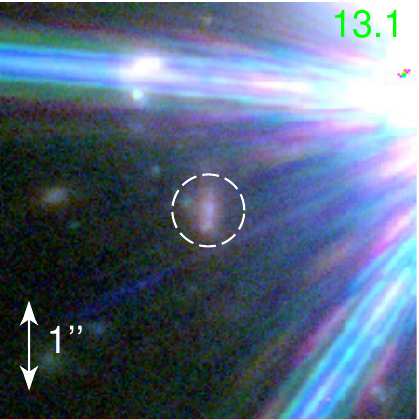}
    \includegraphics[width=0.15\columnwidth]{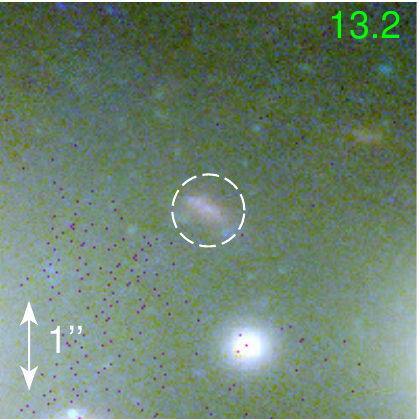}
    \includegraphics[width=0.15\columnwidth]{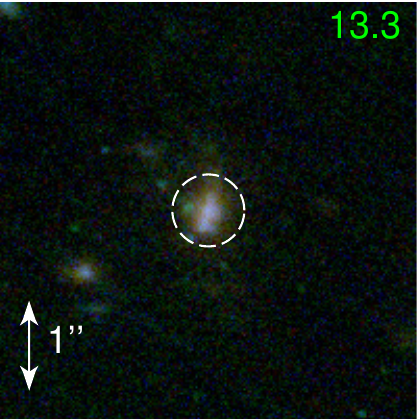}
    \includegraphics[width=0.15\columnwidth]{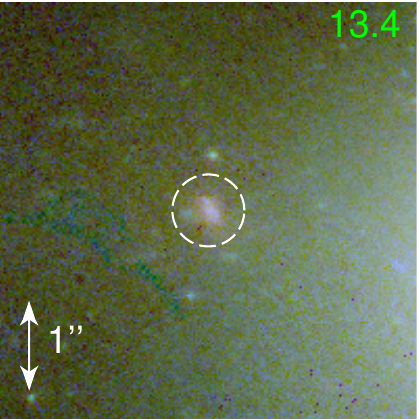}
    \caption{Thumbnails of multiple imaged sources behind \SMACS }
    \label{fig:snippet}
\end{figure}
\begin{figure}
\centering
    \includegraphics[width=0.15\columnwidth]{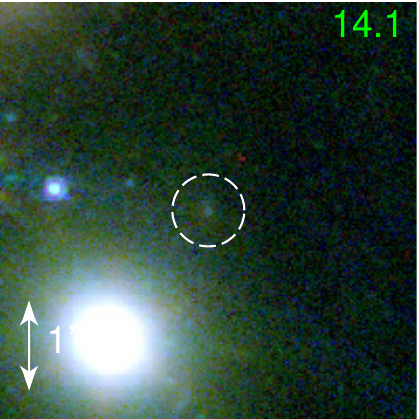}
    \includegraphics[width=0.15\columnwidth]{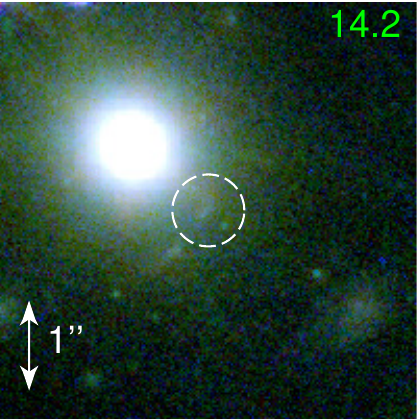}
    \includegraphics[width=0.15\columnwidth]{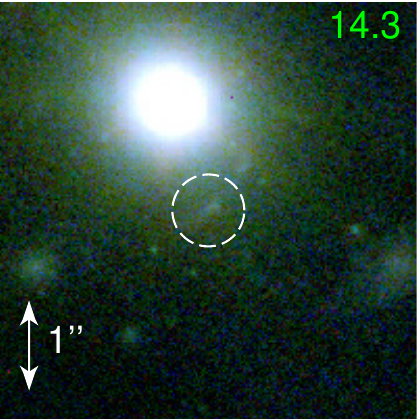}
    \includegraphics[width=0.15\columnwidth]{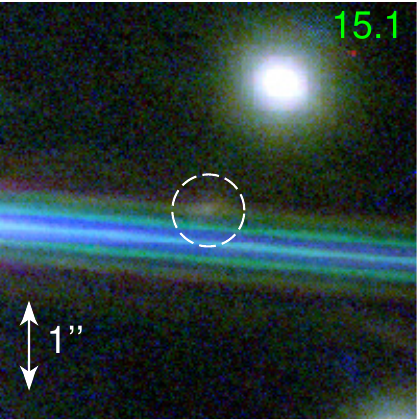}
    \includegraphics[width=0.15\columnwidth]{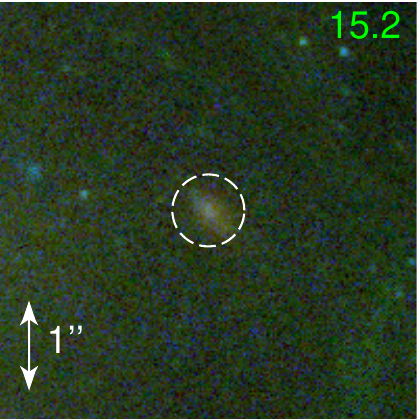}
    \includegraphics[width=0.15\columnwidth]{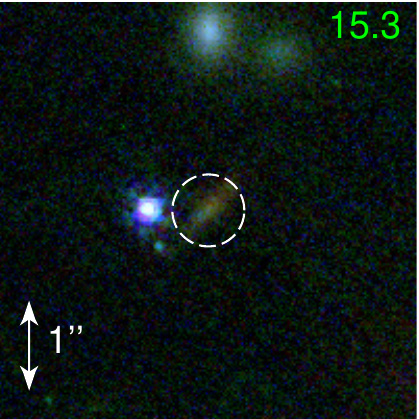}
    \includegraphics[width=0.15\columnwidth]{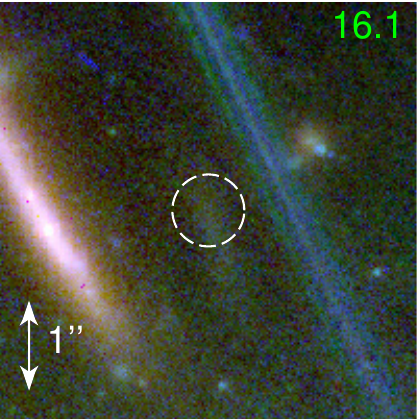}
    \includegraphics[width=0.15\columnwidth]{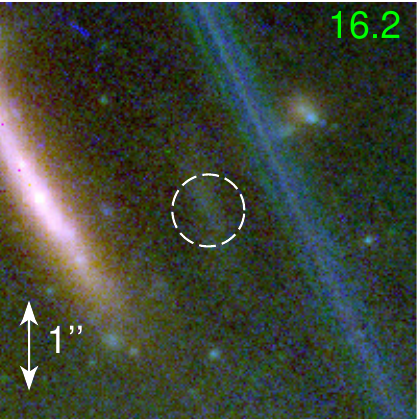}
    \includegraphics[width=0.15\columnwidth]{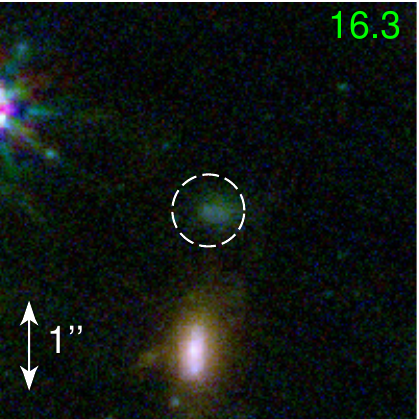}
    \includegraphics[width=0.15\columnwidth]{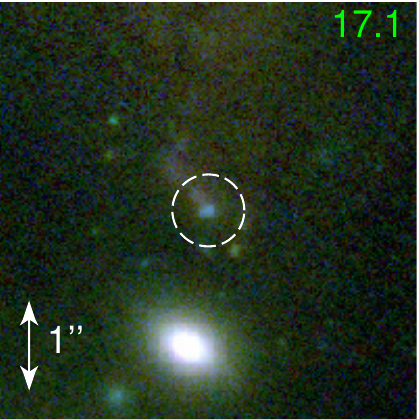}
    \includegraphics[width=0.15\columnwidth]{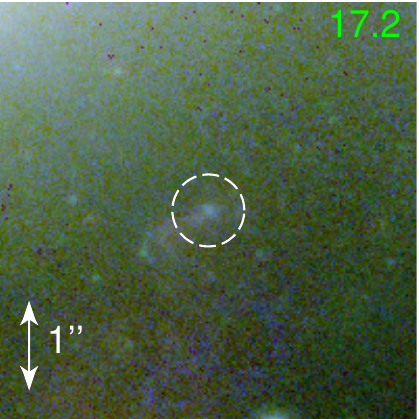}
    \includegraphics[width=0.15\columnwidth]{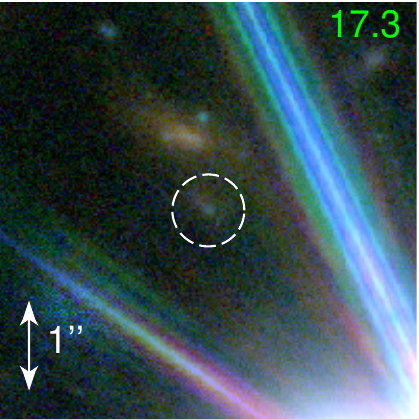}
    \includegraphics[width=0.15\columnwidth]{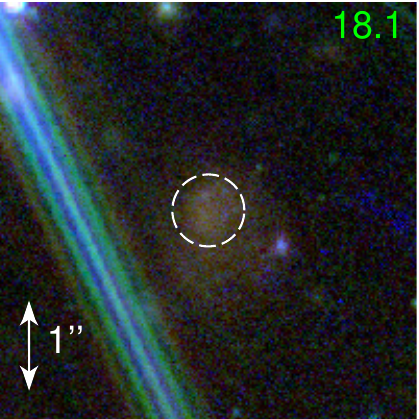}
    \includegraphics[width=0.15\columnwidth]{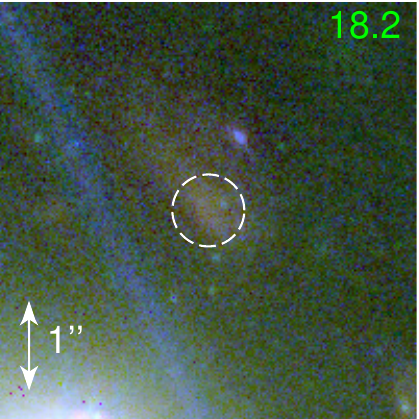}
    \includegraphics[width=0.15\columnwidth]{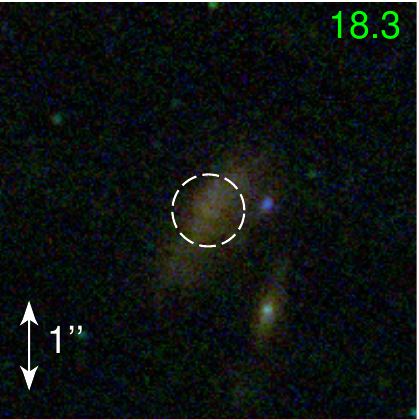}
    \includegraphics[width=0.15\columnwidth]{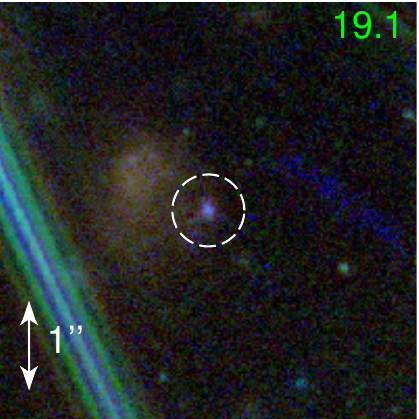}
    \includegraphics[width=0.15\columnwidth]{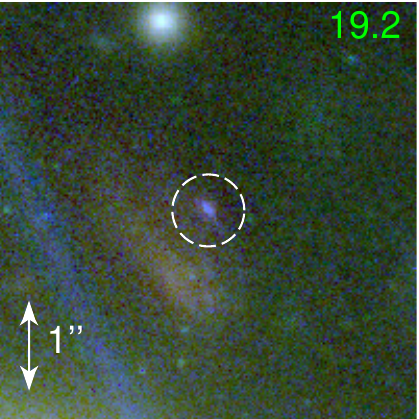}
    \includegraphics[width=0.15\columnwidth]{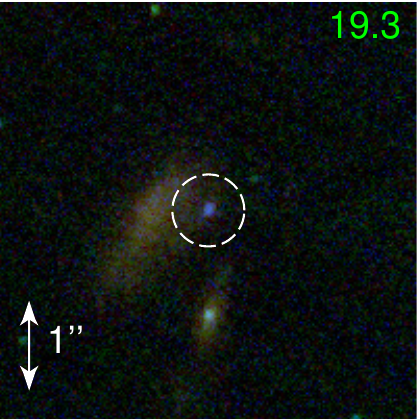}
    \includegraphics[width=0.15\columnwidth]{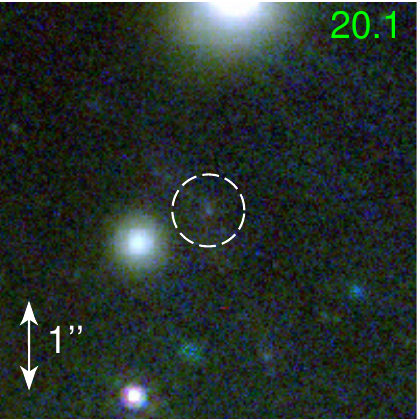}
    \includegraphics[width=0.15\columnwidth]{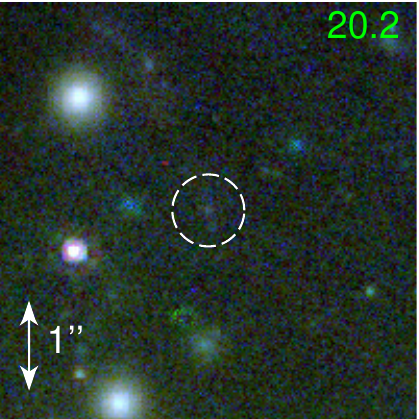}
    \includegraphics[width=0.15\columnwidth]{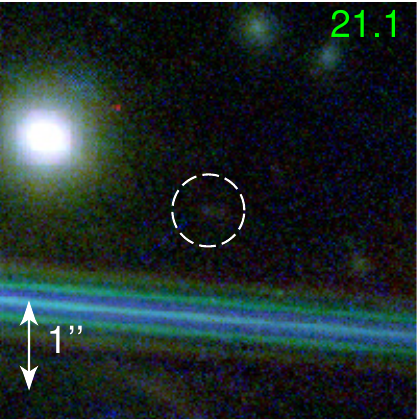}
    \includegraphics[width=0.15\columnwidth]{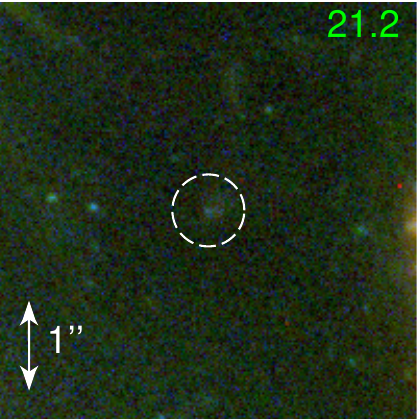}
    \includegraphics[width=0.15\columnwidth]{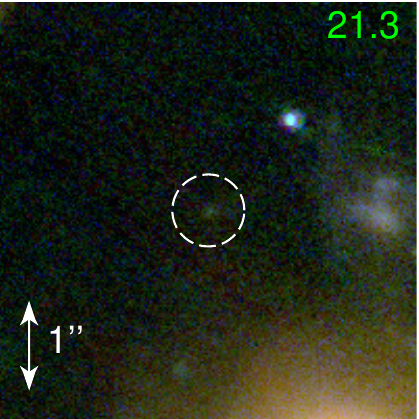}
    \includegraphics[width=0.15\columnwidth]{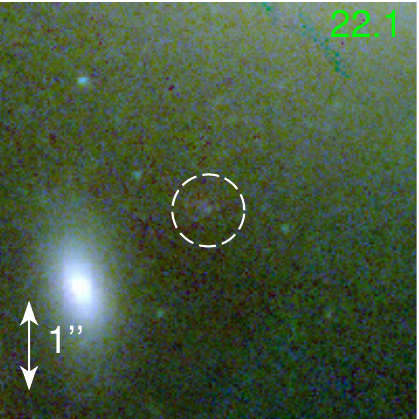}
    \includegraphics[width=0.15\columnwidth]{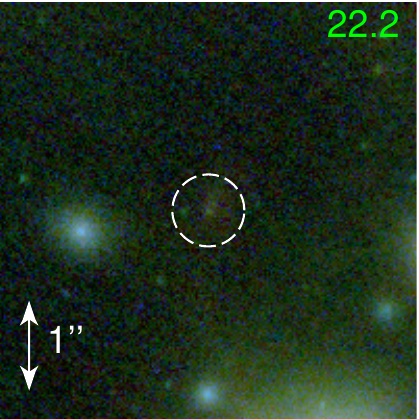}
    \includegraphics[width=0.15\columnwidth]{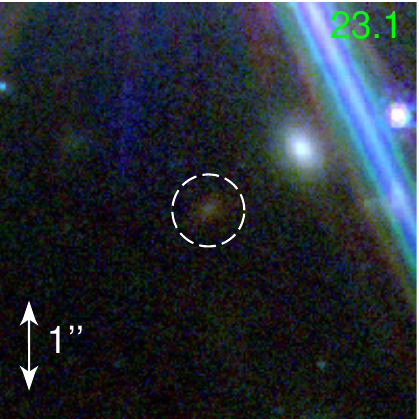}
    \includegraphics[width=0.15\columnwidth]{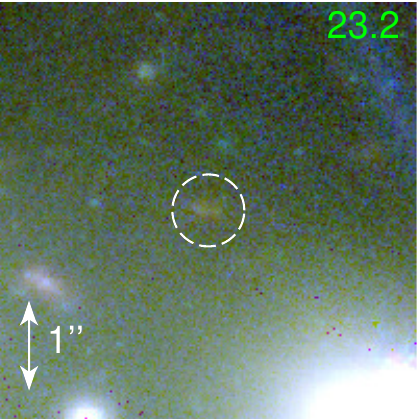}
    \includegraphics[width=0.15\columnwidth]{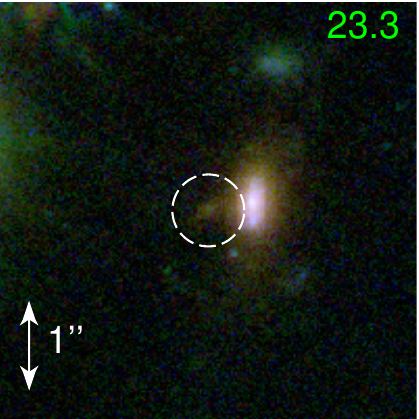}
    \includegraphics[width=0.15\columnwidth]{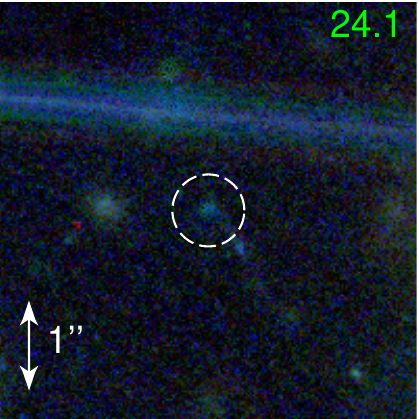}
    \includegraphics[width=0.15\columnwidth]{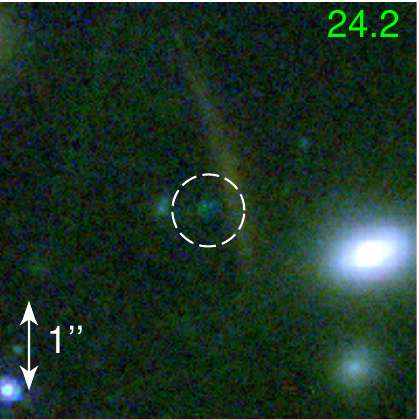}
    \includegraphics[width=0.15\columnwidth]{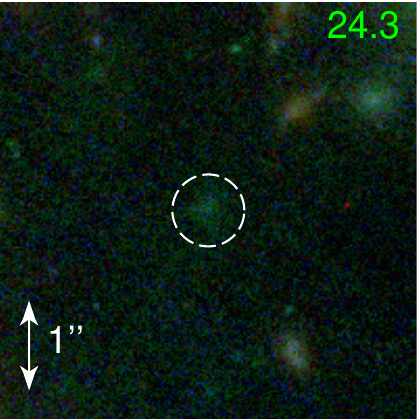}
    \includegraphics[width=0.15\columnwidth]{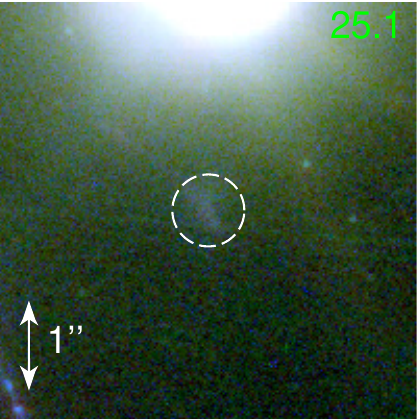}
    \includegraphics[width=0.15\columnwidth]{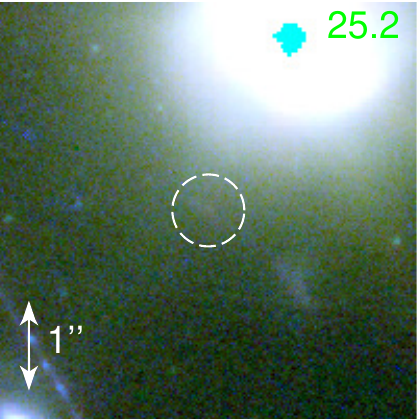}
    \includegraphics[width=0.15\columnwidth]{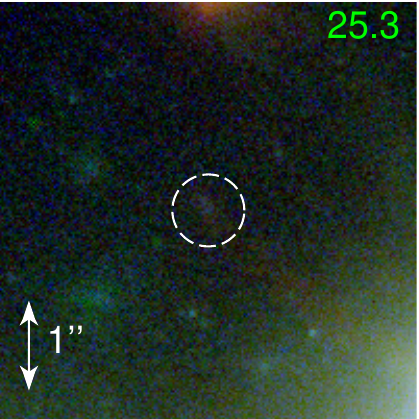}
    \includegraphics[width=0.15\columnwidth]{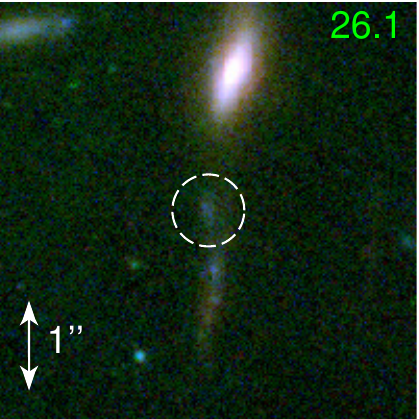}
    \includegraphics[width=0.15\columnwidth]{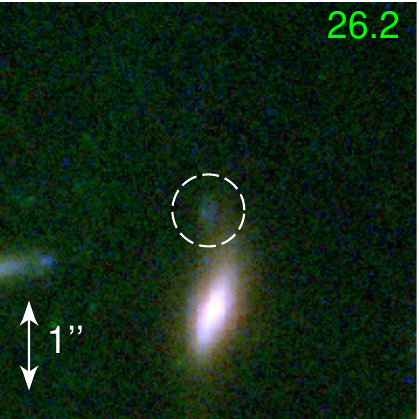}
    \includegraphics[width=0.15\columnwidth]{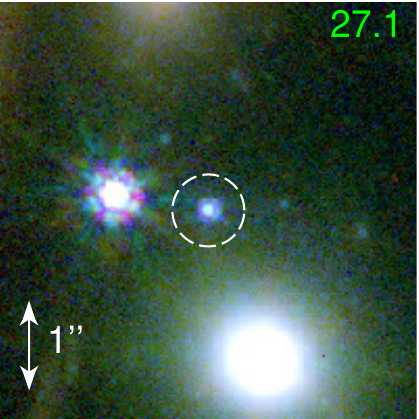}
    \includegraphics[width=0.15\columnwidth]{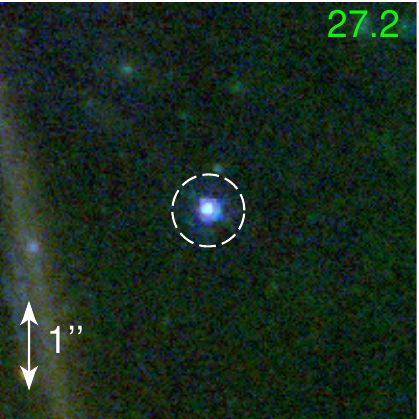}
    \includegraphics[width=0.15\columnwidth]{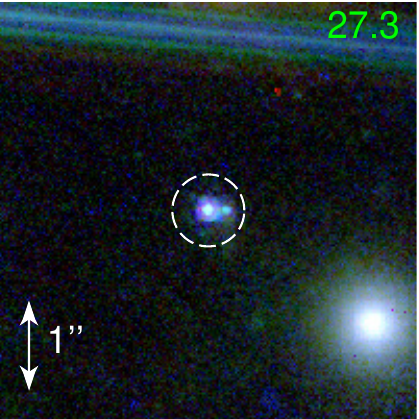}
    \includegraphics[width=0.15\columnwidth]{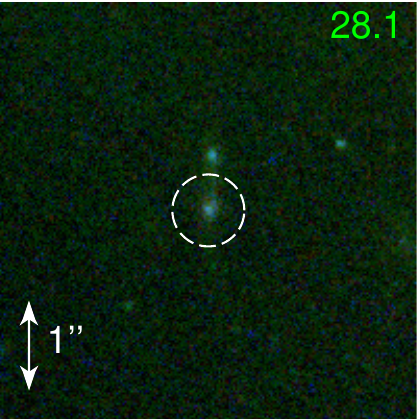}
    \includegraphics[width=0.15\columnwidth]{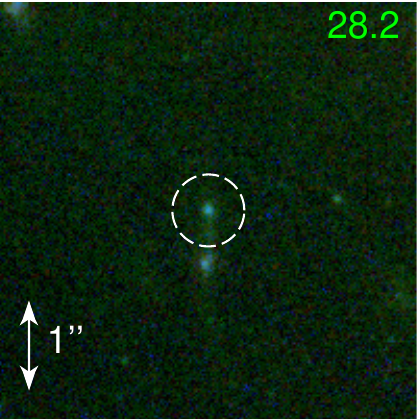}
    \includegraphics[width=0.15\columnwidth]{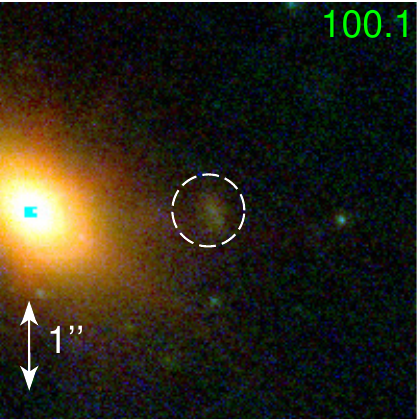}
    \includegraphics[width=0.15\columnwidth]{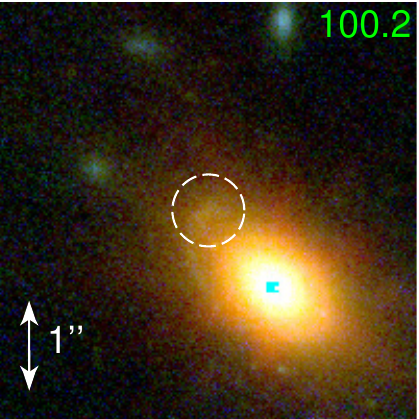}
    \includegraphics[width=0.15\columnwidth]{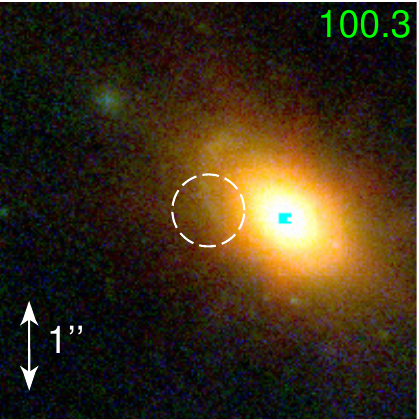}
    \includegraphics[width=0.15\columnwidth]{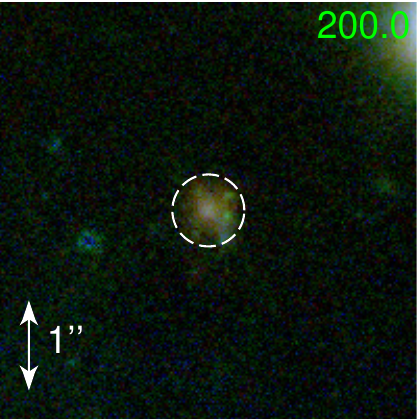}
    \caption{Continuing figure}
    \label{fig:snippet}
\end{figure}

%\end{comment}

%If you want to present additional material which would interrupt the flow of the main paper,it can be placed in an Appendix which appears after the list of references.

%%%%%%%%%%%%%%%%%%%%%%%%%%%%%%%%%%%%%%%%%%%%%%%%%%

\end{document}